\documentclass[aps,pra,a4paper,nofootinbib,superscriptaddress,numbers,longbibliography,showpacs,showkeys,floatfix,dvipsnames,nobalancelastpage,preprint,reprint]{revtex4-1}
\usepackage{amsfonts,amssymb,amsmath}
\usepackage{hyperref,color,graphicx}
\usepackage[separate-uncertainty=true]{siunitx}

\DeclareSIUnit\gauss{G}

\usepackage[english]{babel}
\usepackage[utf8]{inputenc}
\usepackage[T1]{fontenc}
\usepackage{bm}
\usepackage{wasysym}

\usepackage{xspace}

\newcommand{\coloneqq}{\mathrel {\mathop : } \mathrel {\mkern -1.2mu}=}

\DeclareMathOperator{\arccsch}{arccsch}

\def\bra#1{\left\langle#1\right|}
\def\ket#1{\left|#1\right\rangle}
\def\braket#1#2{\left\langle#1\kern-2pt\right.\left|#2\right\rangle}
\def\braOpket#1#2#3{\left\langle#1\right|#2\left|#3\right\rangle}

\newif\ifusebibfile
\usebibfiletrue

\makeatletter \newcommand{\globallabel}[1]{%
    \protected@edef\@currentlabel{\theparentequation}\label{#1}%
}
\makeatother

\iftrue
\addto\captionsenglish{\renewcommand{\figurename}{Fig.}}
\newcommand{\subfiglabel}[1]{\textbf{\MakeUppercase{#1}}}
\newcommand{\figref}[2]{\hyperref[#1]{\ref{#1}{\MakeUppercase{#2}}}}
\makeatletter
\def\fnum@figure{{\bfseries \figurename~\thefigure}}
\def\@caption@fignum@sep{{\bfseries . }}
\makeatother

\def\theTitle{Demonstration of quantum brachistochrones between distant states of an atom}
\fi

\newcommand{\encapsulateMath}[1]{\raisebox{0pt}[0pt][0pt]{#1}}

\hypersetup{
 pdftitle = {\theTitle},
 pdfauthor = {Manolo R.~Lam, Natalie Peter, Thorsten Groh, Carsten Robens, Wolfgang Alt, Dieter Meschede, Antonio Negretti, Simone Montangero, Tommaso Calarco, Andrea Alberti},
	colorlinks=true,linkcolor=blue,citecolor=blue,filecolor=blue,urlcolor=blue,pdfpagemode=UseNone,pdfstartview={XYZ null null 1.00}
}

\renewcommand\textemdash{\leavevmode\unskip\kern0.8pt---\kern1pt\ignorespaces}

\makeatletter
\let\old@Section@Cmd=\section
\def\end@of@sec@tit{.\leavevmode\unskip\kern0.8pt\rule[0.19\baselineskip]{8pt}{0.4pt}\kern1pt}
\def\mysection{%
	\def\reserved@stsec##1{\@startsection{section}{1}{\parindent}{\z@}{0em}{\normalfont\normalsize\itshape}*[##1]{##1\end@of@sec@tit}}%
	\@ifstar{%
		\reserved@stsec%
	}{%
		\reserved@stsec%
	}%
}
\let\section=\mysection
\makeatother

\newcommand{\MT}{Mandelstam-Tamm\xspace}
\long\def\eqref#1{\textup{Eq.~(\ref{#1})}}
\def\supmat{Supp.\ Mat.}

\begin{document}	

\selectlanguage{english}

\title{\theTitle{}}

\newcommand{\affA}{Institut f\"ur Angewandte Physik, Universit\"at Bonn, 53115 Bonn, Germany}
\newcommand{\affB}{MIT-Harvard Center for Ultracold Atoms, Research Laboratory of Electronics, and Department of Physics, Massachusetts Institute of Technology, Cambridge, Massachusetts 02139, USA}
\newcommand{\affC}{Zentrum für Optische Quantentechnologien, Fachbereich Physik, and The Hamburg Centre for Ultrafast Imaging, Universität Hamburg, 22761 Hamburg, Germany}
\newcommand{\affD}{Forschungszentrum Jülich, 52428 Jülich, and Universität zu Köln, 50937  Köln, Germany}
\newcommand{\affE}{Dipartimento di Fisica e Astronomia ``G.~Galilei,'' Università degli Studi di Padova, and Istituto Nazionale di Fisica Nucleare, 35131 Padova, Italy}

\author{Manolo R.~Lam}\affiliation{\affA}
\author{Natalie Peter}\affiliation{\affA}
\author{Thorsten Groh}\affiliation{\affA}
\author{Wolfgang Alt}\affiliation{\affA}
\author{Carsten Robens}\affiliation{\affA}\affiliation{\affB}
\author{Dieter~Meschede}\affiliation{\affA}
\author{Antonio Negretti}\affiliation{\affC}
\author{Simone Montangero}\affiliation{\affE}
\author{Tommaso Calarco}\affiliation{\affD}
\author{Andrea Alberti}\affiliation{\affA}
\email{alberti@iap.uni-bonn.de}

\begin{abstract}
Transforming an initial quantum state into a target state through the fastest possible route \textemdash a quantum brachistochrone \textemdash is a fundamental challenge for many technologies based on quantum mechanics.
Here, we demonstrate fast coherent transport of an atomic wave packet over a distance of $\num{15}$ times its size
\textemdash a paradigmatic case of quantum processes where the target state cannot be reached through a local transformation.
Our measurements of the transport fidelity reveal the existence of a minimum duration \textemdash a quantum speed limit \textemdash for the coherent splitting and recombination of matter waves.
We obtain physical insight into this limit by relying on a geometric interpretation of quantum state dynamics.
These results shed light upon a fundamental limit of quantum state dynamics and are expected to find  relevant applications in quantum sensing and quantum computing.

\end{abstract}

\makeatletter
\let\originalmaketitle=\maketitle
\def\maketitle{%
\@author@finish%
\title@column%
\titleblock@produce%
\suppressfloats[t]%
\let\@AF@join\@AF@join@error%
\titlepage@sw{\vfil\clearpage}{}%
}

\makeatother

\maketitle

\section{Introduction}

How fast can a quantum process be?
Previous efforts to answer this question have resulted in fundamental insights into quantum state dynamics \cite{Mandelstam:1945,Wootters:1981,Bhattacharyya:1983,Anandan:1990,Margolus:1998,Rabitz:2004,Carlini:2006,Bender:2007a,Levitin:2009,Taddei:2013,Deffner:2013a,Lloyd:2014,Pires:2016,Funo:2017,Shanahan:2018,Okuyama:2018,Heck:2018,Bukov:2019}, and shed light onto the ultimate physical limits to the rate of information processing \cite{Bekenstein:1981,Levitin:1982,Lloyd:2000}.
Speeding up the dynamics of a quantum process is also key to advance quantum technologies \cite{Kielpinski:2002,Home:2009,Karski:2009a}, because faster processes can help us outrun detrimental decoherence mechanisms, and so boost the number of high-fidelity operations executed within the system's coherence time \cite{Campbell:2010,Frank:2014,Schaefer:2018}.

The fact that a minimum time is required to accomplish a physical process has been known since Bernoulli's famous brachistochrone problem \cite{Haws:1995}, long before the advent of quantum physics.
The origin of such a minimum time can be traced back to the maximum rate at which a physical state can change in time,
which is generally determined by the amount of physical resources (energy and degree of control) available to carry out the process.

For quantum processes, a precise formulation of such a speed limit was first derived by Mandelstam and Tamm \cite{Mandelstam:1945}
considering the transformation of a quantum state $\ket{\psi_\text{init}}$ into an orthogonal one $\ket{\psi_\text{target}}$.
They discovered that the duration $\tau_\text{QB}$ of the fastest process \textemdash the quantum brachistochrone \textemdash is bound by the inverse of the energy uncertainty \cite{defDeltaH},
\begin{equation}
	\label{eq:MT_bound}
	\tau_\text{QB} \geq \tau_\text{MT} = \frac{\hbar \pi}{2\Delta E},
\end{equation}
providing a firm basis for Heisenberg's time-energy uncertainty principle \cite{Deffner:2017a}.
Most significantly, the \MT bound shows that the duration of a quantum process cannot vanish, unless infinitely large energy resources can be controlled.
An experimental demonstration of this limit was given in effective two-level systems using ultracold atoms \cite{Bason:2012,Frank:2016} and superconducting transmon circuits \cite{Vepsalainen:2019}.

\section{Quantum brachistochrones between distant states}

\begin{figure*}[t]
	\includegraphics[width=1\textwidth]{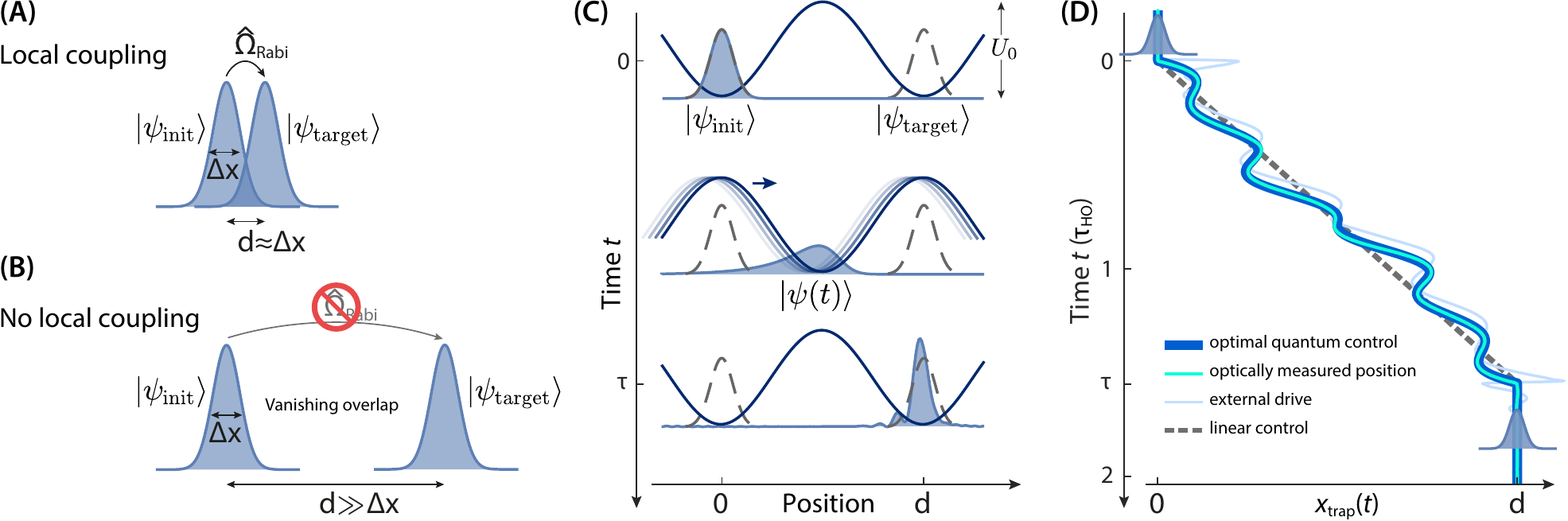}
	\caption{%
			\textbf{Transporting a massive quantum particle to a distant state.}
		(\subfiglabel{a}) 
    Direct local coupling $\hat{\Omega}_{\mathsf{Rabi}}$ between the initial and target state can be realized when the two wave functions have nonzero spatial overlap.
    	(\subfiglabel{B}) Fundamentally, no direct local coupling between the two states can be realized for large separations, $d\gg\Delta x$, suppressing the possibility to attain the \MT bound.
    	(\subfiglabel{C}) Atom transport in an optical conveyor belt (sinusoidal potential curves), depicted at the initial, intermediate and final time of the process.
    The probability distribution of the transported wave packet $\ket{\psi(t)}$ is shown (shaded area), together with that of the initial and target states (dashed lines).
		For illustration purpose, the chosen example shows a wave packet ending up in an excited state, corresponding to a low ($\mathcal{F} \sim 0.5$) transport fidelity.
			(\subfiglabel{D}) Quantum brachistochrone trajectory $x_\text{trap}(t)$ of the optical conveyor belt (dark blue), corresponding to the diamond data point marked by an arrow in Fig.~\ref{Fig:QSL_measurement}.
            							The actual position of the conveyor belt (cyan), measured with $\SI{1}{\angstrom}$ precision by optical interferometry, and the corresponding external drive (light blue), applied to steer the conveyor belt position, are also shown.
	    For comparison, a linear transport ramp (dashed line) is displayed. }
	\label{Fig:Transport_Ramp}
\end{figure*}

Today, it is understood \cite{Brody:2006,Levitin:2009,Hegerfeldt:2013} that the \MT bound in \eqref{eq:MT_bound} can only be saturated (i.e., $\tau_\text{QB}=\tau_\text{MT})$ when the quantum dynamics can be reduced to that of a simple two-level system, i.e., when the target state can be reached directly by a Rabi oscillation (Fig.~\figref{Fig:Transport_Ramp}{a}).
Recently, however, 
the definition of quantum speed limit 
as defined by $\tau_\text{MT}$ has been subject to criticism \cite{Bukov:2019}:
When no direct local coupling between the initial to the final state exists \cite{nonlocal}, as is the case for spatially distant states (Fig.~\figref{Fig:Transport_Ramp}{b}), then the \MT bound does not capture the true quantum speed limit of the process (i.e, \encapsulateMath{$\tau_\text{QB}\gg \tau_\text{MT}$}).

In this work, we give the first experimental demonstration of coherent control of a physical process at its quantum speed limit beyond direct local operations.
Specifically, we consider the problem of transporting a trapped massive
quantum particle to a distant location, separated by about $\num{15}$ times the size of the wave packet, in the minimum possible time;
the initial
and target
states are defined by the ground state of the trap potential centered at the two different locations.
Because of the wide separation between the two states, it is fundamentally impossible for a 
massive quantum particle to reach the target state by a Rabi oscillation.
Any local operator \encapsulateMath{$\hat{\Omega}_\text{Rabi}$} yields in fact a vanishingly small Frank-Condon factor \encapsulateMath{$\braOpket{\psi_\text{init}}{\hat{\Omega}_\text{Rabi}}{\psi_\text{target}}$} (Fig.~\figref{Fig:Transport_Ramp}{b}).

We see that inequality (\ref{eq:MT_bound})
fails to give a meaningful bound on the shortest transport duration $\tau_\text{QB}$ if
we examine its scaling with respect to the transport distance $d$:
While the minimum time $\tau_\text{QB}$ is naturally expected to increase with $d$,
remarkably, the time $\tau_\text{MT}$ exhibits rather the opposite behavior, as it decreases with $d$ (\supmat, Sec.~\ref{app:scalingMandelstamTamm}).
A way out of this conundrum will be discussed below, adopting a geometric point of view on wave packet dynamics.

\begin{figure*}[t]
	\centering
	\includegraphics[width=\textwidth]{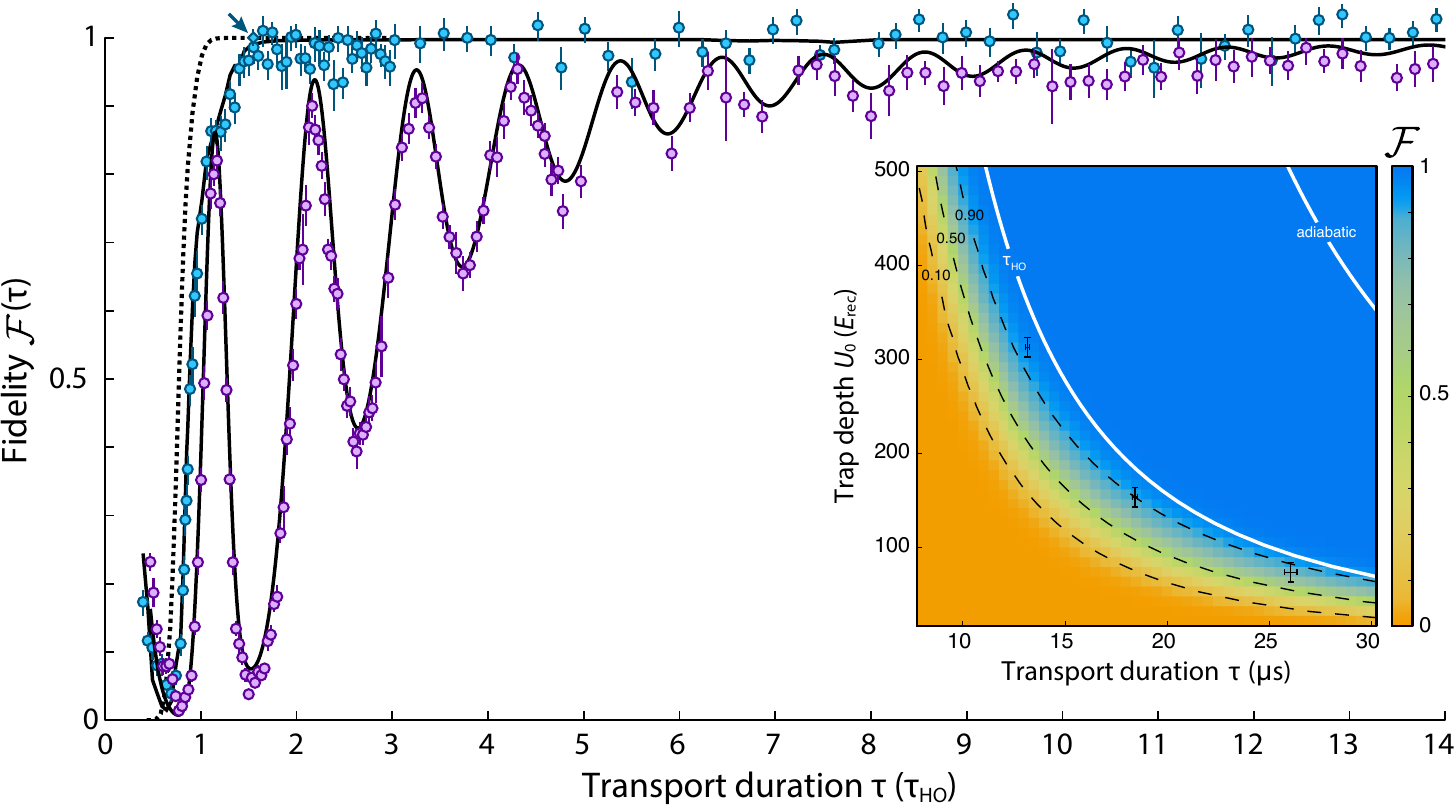}
	\caption[]{\textbf{Revealing the quantum speed limit.}
	The fidelity $\mathcal{F}(\tau)$ of transporting an atom over one lattice site is measured as a function of the transport duration $\tau$, expressed in units of the oscillation period $\tau_\text{HO}{\,\approx\,} \SI{20}{\micro\second}$, for a trap depth $U_0{\,\approx\,} \num{150}\,E_\text{rec}$.
 	Blue points: Optimal quantum control achieves near-unit fidelity for durations above the quantum brachistochrone time $\tau_\text{QB}$ (diamond point marked by an arrow), in the proximity of $\tau_\text{HO}$.
 	Below the quantum brachistochrone time, the fidelity drops rapidly, revealing the existence of a quantum speed limit.
    Purple points: Linear transport ramps achieve sub-optimal fidelity.
 	Black lines: computed fidelity based on numerical simulations of atom transport, assuming a transverse temperature $T_\perp {\,\approx\,} \SI{1}{\micro\kelvin}$ (solid) and a zero-temperature case (dashed).
 		   Inset: The fidelity landscape computed numerically as a function of $U_0$ and $\tau$ for $T_\perp=0$ (colored contour map) and the measured transition points (experimental data) where the fidelity reaches $\mathcal{F}{\,\approx\,} 0.5$.
	White lines: the oscillation period $\tau_\text{HO}$, approximately representing the quantum brachistochrone time $\tau_\text{QB}$, and the adiabatic limit ensuring fidelities $\mathcal{F} {\,>\,} \num{0.9}$.
	   Error bars represent one standard deviation.
		}
	\label{Fig:QSL_measurement}
\end{figure*}

\section{Fast atom transport in optical conveyor belts}

As of yet, transport experiments have been performed with trapped ions and ultracold atoms in the nearly harmonic low-energy portion of the trap potential \cite{Couvert:2008,Bowler:2012,Walther:2012,Alonso:2016,Ness:2018}, where fast, high-fidelity transport is enabled by effective protocols  \cite{Murphy:2009a,Torrontegui:2011} known as shortcuts to adiabaticity.
In order to reach the quantum speed limit, however, excitations of the wave packet beyond the low-energy range of the trap potential must be controlled, requiring precise knowledge of the full potential.
For this purpose, we employ a one-dimensional optical lattice to transport neutral atoms along its axis (Fig.~\figref{Fig:Transport_Ramp}{c}), acting like a conveyor belt \cite{Miroshnychenko:2006b}.
The resulting sinusoidal potential is inherently well defined over all spatial regions from trough to crest, since it is created by optical interference of two counterpropagating laser beams (wavelength $\lambda\approx\SI{866}{\nano\meter}$, lattice constant $\lambda/2$). We choose the trap depth $U_0$
of the order of $100\,E_\text{rec}$ in order to suppress tunneling of the initial state to adjacent sites; $E_\text{rec} = (2\pi\hbar)^2/(2m\lambda^2)$ is the recoil energy of an atom of mass $m$.
We also maintain $U_0$ constant during the whole transport process to explore the scenario where the energy available to control a physical process is fundamentally limited;
in fact, for an infinitely deep potential, no speed limit exists \cite{Murphy:2009a} in non-relativistic quantum mechanics.

All transport experiments begin
by preparing the matter wave of a ${}^{133}$Cs atom into the motional ground state $\ket{\psi_\text{init}}$ of one of the sites of the optical conveyor belt (\supmat, Sec.~\ref{app:ExperimentalApparatus}), which is initially held at rest.
Subsequently, we displace the conveyor belt within a given time $\tau$ to the desired target location following a chosen trajectory $x_\text{trap}(t)$ as a function of time $t$ (Fig.~\figref{Fig:Transport_Ramp}{d}).
The target location is chosen to be one lattice site away ($d = \lambda/2$) from the initial location, corresponding to $\num{15}$ times the initial size $\Delta x$ of the wave packet.
While the atomic wave packet is highly excited during transport, it ideally ends up in the ground state of the displaced potential, $\ket{\psi_\text{target}}$, once the optical conveyor belt is brought back to rest.
We conclude the experiments by measuring (\supmat, Sec.~\ref{app:detection_fidelity}) the fidelity of the transport process,
\begin{equation}
	\label{eq:fidelity_def}
    \mathcal{F}(\tau)=\left|\braket{\psi_\text{target}}{\psi(\tau)}\right|^2,
\end{equation}
quantifying the probability of occupying $\ket{\psi_\text{target}}$.

In the experiments, we control the position $x_\text{trap}(t)$ of the optical conveyor belt with high precision using a fast polarization synthesizer \cite{Robens:2018}, reducing the position noise $\delta x$ 
to much less than the size of the wave packet $\Delta x$ ($\delta x \:{\approx}\: \SI{0.1}{\nano\meter} \:{\ll}\: \Delta x \:{\approx}\: \SI{25}{\nano\meter}$).
We additionally suppress systematic distortions from the desired trajectory, which are caused by the finite bandwidth ($\approx\SI{1}{\mega\hertz}$) and nonlinearities of the control setup, by suitably oversteering the applied external drive signal (Fig.~\figref{Fig:Transport_Ramp}{d});
the drive signal is derived by applying a combination of deconvolution and iterative compensation techniques (\supmat, Sec.~\ref{app:deconvolution}).
Time-resolved measurements of $x_\text{trap}(t)$, carried out by on-site laser interferometry (\supmat, Sec.~\ref{app:opticalMeasurement}), reveal a nearly perfect agreement between the actual trajectory of the conveyor belt and the targeted one (Fig.~\figref{Fig:Transport_Ramp}{d}), with peak-to-peak discrepancies less than $\SI{10}{\nano\meter}$.

\section{Optimal transport solutions}

We steer the conveyor belt along trajectories that are specially chosen to maximize the transport fidelity, thus realizing feed-forward quantum control \cite{Guery-Odelin:2019}.
For a fixed duration $\tau$, to obtain an optimal trajectory $x_\text{trap}(t)$, we take the solution for the corresponding classical problem (\supmat, Sec.~\ref{app:ClassicalLimitAndInitialGuess}), and subsequently apply optimal quantum control methods \cite{Caneva:2009,Koch:2016,Chen:2015b} to maximize $\mathcal{F}(\tau)$, relying on numerical simulations of atom transport.
In the optimization problem, two constraints must be fulfilled:
(1) $x_\text{trap}(t\,{\leq}\,0)=0$ and $x_\text{trap}(t\,{\geq}\,\tau)=d$; (2) the Fourier spectrum of $x_\text{trap}(t)$ is limited to within the control setup bandwidth in order to ensure that $x_\text{trap}(t)$ is faithfully reproduced in the experiments.
Importantly, the bandwidth constraint does not significantly affect the maximum attainable fidelity, provided that the control setup bandwidth exceeds $U_0/(2\pi\hbar)$,
 which is the case here (\supmat, Sec.~\ref{app:OptimalControlOfAtomTransport}).

The resulting optimal trajectories (Fig.~\figref{Fig:Transport_Ramp}{d}) exhibit a rather wiggling behavior, which is key to control excitations during transport.
Disregarding the fast wiggles, the remaining behavior of $x_\text{trap}(t)$ is reminiscent of a constantly accelerated and decelerated trajectory for the first and second half of the transport duration.
In addition, optimal trajectories notably start and finish with swift displacements, which are favorable to place the atomic wave packet where the trap potential is steep (\supmat, Sec.~\ref{app:ClassicalLimitAndInitialGuess}).

\section{Revealing the quantum speed limit}

Our measurements of the transport fidelity (Fig.~\ref{Fig:QSL_measurement}) demonstrate that optimal quantum-control solutions accomplish $\mathcal{F}\approx1$ within experimental uncertainty for all transport times greater than $\tau_\text{QB}$, occurring in the proximity of $\tau_\text{HO}$,
the oscillation period in the harmonic approximation of the trap potential (\supmat, Sec.~\ref{app:ClassicalLimitAndInitialGuess}).
Crucially, for times shorter than $\tau_\text{HO}$, the fidelity drops rapidly, revealing the existence of a minimum duration \textemdash a quantum speed limit \textemdash for the transport of matter waves.
To our knowledge, this is the first observation of the quantum speed limit for a multi-level system, where the transition from a quantum-controllable to a quantum-noncontrollable process is sharply resolved by fidelity measurements.
Instead of a sharp transition from perfectly controllable to completely uncontrollable, as in the classical analog problem, we here observe a rapid, yet smooth crossover around $\tau_\text{QB}$ \cite{Sels:2018}.

We obtain insight about $\tau_\text{QB}$ by exploring the fidelity landscape $\mathcal{F}(\tau)$ as a function of $\tau$, for different trap depths $U_0 \approx \{70,150,300\}\,E_\text{rec}$.
By varying the trap depth, we change the number of effectively controlled energy levels ($\num{4},\,\num{6},\,\num{10}$, respectively), for which site-to-site tunneling is negligible over the transport duration $\tau$.
We determine the transition to a quantum noncontrollable process as the transport time at which the measured fidelity drops to $\mathcal{F}(\tau)\approx 0.5$ (inset of Fig.~\ref{Fig:QSL_measurement}).
Our measurements demonstrate that in the range of trap depths explored here, the quantum brachistochrone time $\tau_\text{QB}$ follows approximately $\tau_\text{HO}$.
Atom transport performed in a time close to $\tau_\text{HO}$ is notably much faster than its adiabatic counterpart, which requires on the contrary $\tau {\,\gg\,} \tau_\text{HO}$ (\supmat, Sec.~\ref{app:adiabatic_limit}).
Further insight into the scaling of $\tau_\text{QB}$ is provided in the final discussion, relying on geometric arguments.

To validate our experimental results, we employ numerical simulations of the transport process based on a one-dimensional model of the conveyor belt potential (\supmat, Sec.~\ref{app:simulations}).
A direct comparison of the computed fidelity with the measured $\mathcal{F}(\tau)$ reveals an excellent agreement with the simulations taking into account a thermal distribution in the transverse direction to the optical conveyor belt (Fig.~\ref{Fig:QSL_measurement}).
Relying on the numerical simulations, we are able to explain the rapid drop of fidelity observed when the transport duration is reduced below $\tau_\text{QB}$:
For short durations, high-energy excitations are created above the discrete spectrum of controlled energy levels, leading to a significant probability of tunneling to the neighboring sites and thus to a drop of fidelity.
The occurrence of tunneling is especially evident in the limit of very short durations, $\tau {\,\ll\,} \tau_\text{QB}$.
In this limit, in fact, the optical conveyor belt is displaced so fast that the atom has a considerable probability to remain in the very same state $\ket{\psi_\text{init}}$ where it was initially prepared.
This possibility explains the rise in fidelity for very short times observed in Fig.~\ref{Fig:QSL_measurement};
such an event could be separately detected by resolving the individual lattice sites \cite{Alberti:2016} in addition to measuring the ground state probability.

For comparison, we perform analogous transport experiments applying a simple linear transport ramp (Fig.~\figref{Fig:Transport_Ramp}{d}), corresponding to a bang-bang type of control (\supmat, Sec.~\ref{app:bang-bang}), as opposed to optimal quantum control.
In spite of its simplicity, bang-bang control enables faster-than-adiabatic high-fidelity transport, and finds wide applications in quantum technology \cite{Alonso:2016}.
The measured transport fidelity reveals maxima of $\mathcal{F}(\tau)$ when the transport duration is chosen to be a multiple of the oscillation period $\tau_\text{HO}$ (Fig.~\ref{Fig:QSL_measurement}).
In an ideal harmonic trap, these maxima are expected to reach unit fidelity owing to a perfect refocussing of motional excitations (\supmat, Sec.~\ref{app:bang-bang}).
Our measurements show, however, that such refocusing mechanism is only partially effective, owing to the anharmonicity of the conveyor belt potential.
To reach fidelity values close to unity, long transport times are required, $\tau {\,\gg\,} \tau_\text{HO}$, rendering bang-bang control in anharmonic potentials nearly as ineffective as adiabatic transport.

\begin{figure}
	\centering
	\includegraphics[width=1\columnwidth]{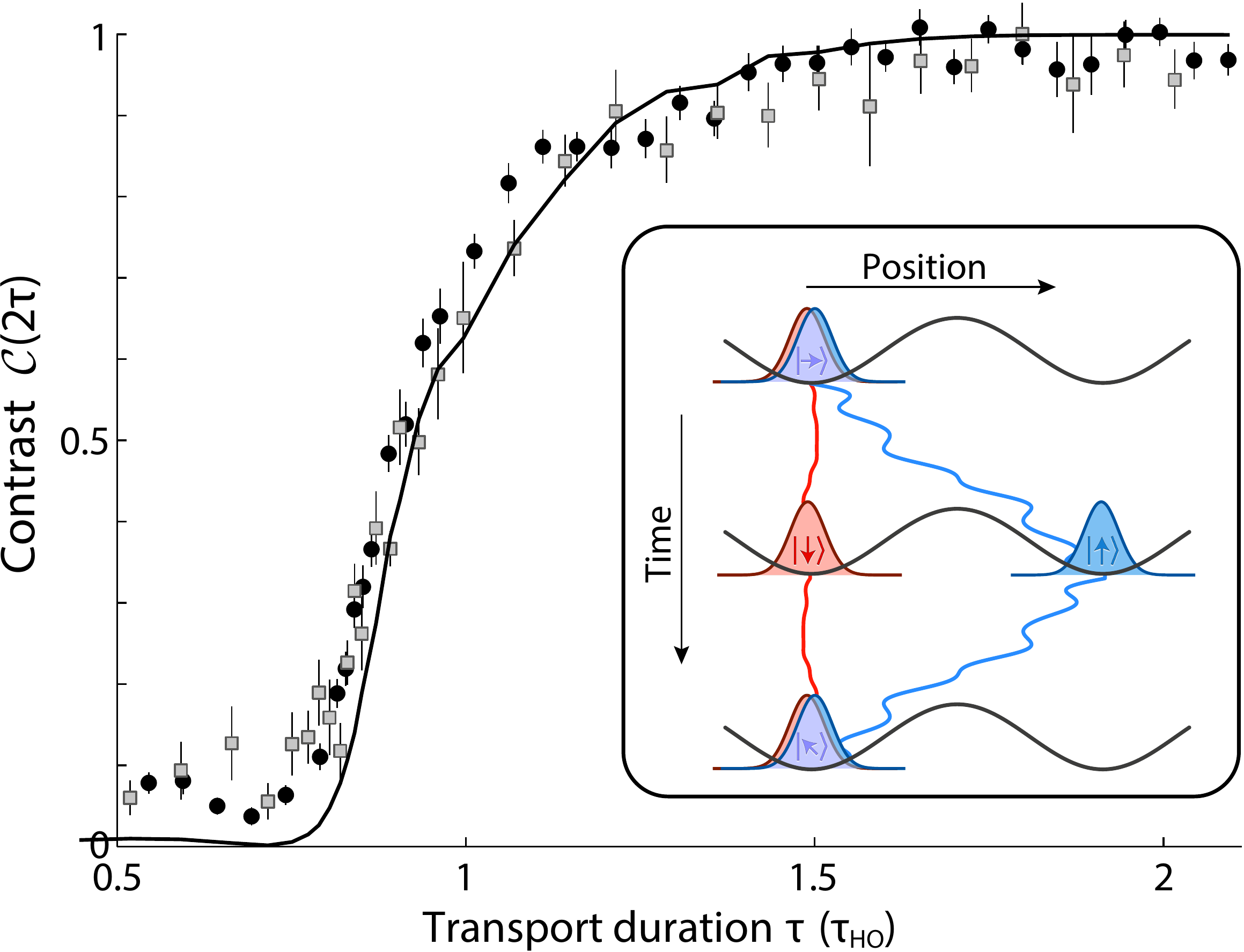}
	\caption{\textbf{Atom interferometry at the quantum speed limit.} 
	Square points: measured contrast $\mathcal{C}(2\tau)$ of the atom interferometer of duration $2\tau$, with the atom in $\ket{\uparrow}$ being transported with optimal quantum control back and forth.
	    Circle points: measured fidelities $\mathcal{F}(\tau)$ from Fig.~\ref{Fig:QSL_measurement}, reproduced here for comparison.
    		Solid line: expected contrast obtained from numerical simulations assuming $T_\perp {\,\approx\,} \SI{1}{\micro\kelvin}$.
		Error bars represent one standard deviation.
	Inset: the blue and red lines describe the movement of the spin-up and spin-down conveyor belts, ensuring that the atom in spin-down state remains effectively at rest, since the two spin-dependent potentials are not fully decoupled (\supmat, Sec.~\ref{app:lattice_potential}). 
}
	\label{Fig:Interferometer}
\end{figure}

\section{Coherent splitting and recombination of matter waves}

To demonstrate that optimal quantum control transport is fully coherent, we conduct a second, closely related experiment, realizing a single-atom Mach-Zehnder interferometer.
To this purpose, we create a copy of the initial atom wave packet with opposite spin direction, realizing a superposition of $\ket{\uparrow}$ and $\ket{\downarrow}$ states, subject to two fully independent, spin-selective optical conveyor belts \cite{Robens:2018}.
Keeping the initial spin-down state at rest, we transport the spin-up state to the next lattice site employing an optimal quantum-control trajectory of duration $\tau$, and bring it back with the same trajectory reversed.
We conclude the interferometer experiments by retrieving the contrast $\mathcal{C}(2\tau)$ of the interference fringe with a Ramsey interrogation scheme \cite{Steffen:2012}.

In analogy to our previous findings, the interferometer measurements reveal a high contrast for transport durations $\tau{\,\gtrsim\,} \tau_\text{HO}$, attesting to the fully coherent nature of the process  (Fig.~\ref{Fig:Interferometer}).
The measured contrast is in fact directly related to the fidelity  $\mathcal{F}_2(2\tau)$ of the process transporting the atomic wave packet back and forth: \encapsulateMath{$\mathcal{C}(2\tau) = \left| \braket{\psi_\text{init}}{\psi(2\tau)}\right| = \sqrt{\mathcal{F}_2(2\tau)}$}.
Moreover, if we make the assumption $\mathcal{F}_2(2\tau) \approx \mathcal{F}(\tau)^2$, we can trace $\mathcal{F}_2(2\tau)$ back to the single transport fidelity.
The direct comparison of the measurements of the fidelity $\mathcal{F}(\tau)$ and contrast $\mathcal{C}(2\tau)$ reveals a striking similarity (Fig.~\ref{Fig:Interferometer}).
This comparison shows the importance of achieving high-fidelity transport operations for fully coherent quantum processes involving superposition of states.

\vspace{3mm}

\section{Interpretation and physical insight}
\label{sec:InterpretationAndInsight}

A basic interpretation of the quantum brachistochrone time $\tau_\text{QB}$ observed in the experiments is provided by the analog classical problem.
There, the fastest process is realized when the particle is maximally accelerated for half of the time and then decelerated for the other half, with its position being centered at the points of steepest potential.
This protocol results in the classical brachistochrone time \encapsulateMath{$\tau_\text{CB} = \tau_\text{HO} \sqrt{2\,n/\pi}$}, where $n$ represents the transport distance $d$ expressed in units of the lattice constant $\lambda/2$ (\supmat, Sec.~\ref{app:ClassicalLimitAndInitialGuess}).
When transporting a quantum particle, however, extra control is necessary to
prevent too large spreading of the wave packet in the anharmonic potential \cite{Sels:2018}, in particular when the wave packet approaches the points of steepest potential, where the trap loses its ability to confine.
This additional requirement translates in a longer time to achieve a near-unit fidelity ($\tau_\text{QB} {\,>\,} \tau_\text{CB}$), yielding a lower bound on the quantum brachistochrone time,
\begin{equation}
    \label{eq:classical_bound}
    \tau_\text{QB} > \tau_\text{HO} \sqrt{2\,n/\pi},
\end{equation}
where $n{\,=\,}1$ is the case chosen for the experiments in this work.
Comparing this bound to the measured data, which show near-unit fidelity for durations above $\tau_\text{HO}$, validates the finding that the transport of atoms in our experiments attains the quantum speed limit.

Can the same bound in \eqref{eq:classical_bound} be obtained from quantum mechanical principles?
As we argued earlier, this question cannot be answered based on the \MT bound.
Instead, we consider the quantum state evolution from a geometric point of view, as proposed by Anandan and Aharonov \cite{Anandan:1990}.
 They prove that for every quantum process of duration $\tau$, the average energy uncertainty $\Delta E$ \cite{defDeltaH} is related to the geometric path length of the time-evolved state $\ket{\psi(t)}$,
\begin{equation}
    \label{eq:AA_exp}
    \ell = \int_0^{\tau}\mathrm{d}s_\text{FS} =  \Delta E \, \tau/\hbar,
\end{equation}
measured by the Fubini-Study metric in the Hilbert space of quantum states, $\mathrm{d}s_\text{FS}^2 = 1-\left|\braket{\psi(t+ \mathrm{d} t)}{\psi(t)}\right|^2$ \cite{FubiniStudy}.
Applying this relation to a quantum brachistochrone process, we directly obtain a lower bound on the quantum brachistochrone time, $\tau_\text{QB} > \hbar\,\ell_\text{QB}/{\Delta E_\text{upper}}$, provided that (I) the path length $\ell_\text{QB}$ of the process is known, and (II) an upper bound $\Delta E_\text{upper}$ on the average energy uncertainty can be provided.

To produce (I) and (II), we assume that at the quantum speed limit the wave packet is steadily accelerated in the first half and steadily decelerated in the second half, with its shape maintained close to that of a coherent state.
Concerning point (I), under this assumption we can estimate the path length of the quantum brachistochrone process as the product of two factors (\supmat, Sec.~\ref{app:estimation_DeltaE}),
\begin{equation}
	\label{eq:approx_ell}
	\ell_\text{QB} \approx  \frac{d}{2\Delta x} \,f\!\left(\frac{\tau_\text{HO}}{\pi\tau_\text{QB}}\right),
\end{equation}
where $f(\xi) = \sqrt{1+\xi^2} + \xi^2 \arccsch(\xi)$ is a monotonically increasing function greater than $\num{1}$ for positive arguments.
Notably, the first factor in \eqref{eq:approx_ell}
coincides with the distance between the initial and final states as measured by the quantum geometric tensor (\supmat, Sec.~\ref{app:estimation_DeltaE}),
\begin{equation}
	\label{eq:geo_path_len}
\ell_\text{QGT}=\frac{d}{2\Delta x}.
\end{equation}
In contrast to $\ell_\text{QB}$ in \eqref{eq:approx_ell}, $\ell_\text{QGT}$ is a purely geometric quantity independent of the 
dynamics of the process, since it represents the shortest path length as measured by the Fubini-Study metric in the restricted manifold of
static states that are reachable via an adiabatic transformation of the control parameter $x_\text{trap}$ (\supmat, Sec.~\ref{app:estimation_DeltaE}).
Equation (\ref{eq:approx_ell}) shows that $\ell_\text{QB}$ is larger than $\ell_\text{QGT}$.
This finding is in line with the conjecture put forward in Ref.~\cite{Bukov:2019}
that $\ell_\text{QGT}$ is a lower bound on the path length $\ell$ of those processes that are realizable with the control parameters available (in this work, $x_\text{trap}$),
\begin{equation}
	\label{eq:conjecture}
 	\ell \geq\ell_\text{QGT}.
 \end{equation}

The two factors in \eqref{eq:approx_ell} can thus be interpreted as follows: The first factor $\ell_\text{QGT}$ is a measure of the change of $\ket{\psi(t)}$ when its position is moved across a distance $d$, which can be loosely understood as the number of local transformations necessary to carry out the transport process.
The second factor $f$ instead carries information about the dynamics, reflecting the change of $\ket{\psi(t)}$ when the momentum is varied during transport.
Numerical simulations show that Eq.~(\ref{eq:approx_ell}) approximates the actual $\ell_\text{QB}$ to within a few percent.

Concerning point (II), the determination of an upper bound on $\Delta E$, we bound from above the potential contribution to the instantaneous energy uncertainty $\Delta E(t)$ by assuming the wave packet of size $\Delta x$ to be
positioned where the trap potential is steepest, at $\pm \lambda/8$ from the center of the site (\supmat, Sec.~\ref{app:estimation_DeltaE}).
By averaging over time \cite{defDeltaH}, we thus find an upper bound $\Delta E_\text{upper}$ on $\Delta E$, which remarkably can be expressed in the form
\begin{equation}
	\label{eq:DeltaE_bound}
	\Delta E 
		 < \Delta E_\text{upper}= \ell_\text{QGT}\,f\!\left(\frac{\tau_\text{QB}}{2\,n\,\tau_\text{HO}}\right)\frac{\hbar}{\tau_\text{QB}},
	\end{equation}
where $\ell_\text{QGT}$ originates from the kinetic contribution to $\Delta E(t)$, whereas the second factor $f$ stems from the trap potential contribution.
Combining Eqs.~(\ref{eq:approx_ell}) and (\ref{eq:DeltaE_bound}) in the Anandan-Aharonov relation (\ref{eq:AA_exp}), we obtain
\begin{equation}
	\tau_\text{QB} = \frac{\ell_\text{QB}}{\Delta E/\hbar} > \tau_\text{QB}\: f\!\left(\frac{\tau_\text{HO}}{\pi\tau_\text{QB}}\right)\!\bigg/\!f\!\left(\frac{\tau_\text{QB}}{2\,n\,\tau_\text{HO}}\right),
\end{equation}
which, because of the monotonicity of $f$, directly translates in inequality (\ref{eq:classical_bound}), thus providing a positive answer to the question raised in the beginning.
This result is consistent with the recent findings that the quantum speed limit is not a purely quantum phenomenon, but a universalproperty of the dynamics of physical states in Hilbert space \cite{Shanahan:2018,Okuyama:2018}.

The conjectured bound on the path length, \eqref{eq:conjecture},
alone is not sufficient to yield a bound on $\tau_\text{QB}$, since it does not take into account the dynamical contribution, represented by $f$ in Eq.~(\ref{eq:approx_ell}).
Even so, this bound in \eqref{eq:conjecture} allows us to obtain novel insights applicable to any transport process
that connects spatially distant states.
In fact, using this bound, we find that $\ell$ is not just longer, but significantly longer than the geodesic \textemdash the  shortest possible path as defined by the Fubini-Study metric \textemdash
connecting the initial to the target state.
The reason is that the geodesic coincides \cite{Wootters:1981} with the path in Hilbert space
traced by a Rabi oscillation ($\Omega{\,=\,}\pi/\tau$),
\begin{equation}
	\label{eq:geodesic}
    \ket{\psi(t)} = \cos(\Omega t) \ket{\psi_\text{init}} + \sin(\Omega t) \ket{\psi_\text{target}}.
\end{equation}
whose length is $\ell_\text{geo} = \arccos(\left|\braket{\psi_\text{target}}{\psi_\text{init}} \right|)$.
Importantly, $\ell_\text{geo}$ levels off to $\pi/2$ for orthogonal states, regardless of the distance $d$ separating the two states in real space, thus yielding $\ell\gg \ell_\text{geo}$ for $d{\,\gg\,}\Delta x$.
The atom, in contrast, cannot evolve as in \eqref{eq:geodesic} because,
as a massive particle, it cannot disappear from the initial location while reappearing at the target location \cite{quasiparticles},
but must take a different much longer path.

The geometric relation just obtained, $\ell{\,\gg\,}\ell_\text{geo}$, is the fundamental reason why the \MT inequality falls short of giving a meaningful bound on the quantum brachistochrone duration, $\tau_\text{QB}\gg \tau_\text{MT}$.
In fact, applying the Anandan-Aharonov relation (\ref{eq:AA_exp}) to the quantum brachistochrone process, we directly obtain $\tau_\text{QB} = \hbar\,\ell_\text{QB}/\Delta E \gg \hbar\,\ell_\text{geo}/\Delta E = \tau_\text{MT}$, where $\ell_\text{QB}$ represents the path length of the process, while $\tau_\text{MT}$ designates here the \MT bound generalized \cite{Deffner:2017a} to the case of not necessarily orthogonal states.

\section{Conclusions and outlook}

In this work, we have experimentally demonstrated high-fidelity transport of matter waves connecting spatially distant states in the shortest possible time.
By splitting and recombining atomic matter waves, we showed that coherent quantum control is preserved at the quantum speed limit.
By using geometric arguments, we showed how our transport experiments connecting distant states
go beyond the quantum-speed-limit paradigm developed for single qubits and complex systems that can be effectively reduced to a two-level system \cite{Caneva:2011}, where the \MT bound is known to provide a meaningful lower bound on the shortest duration $\tau_\text{QB}$.
This work focused on a transport distance equal to one lattice site, which is the most relevant case for quantum walks \cite{Karski:2009a}.
It remains for future work to demonstrate quantum brachistochrone for long baseline interferometry, which is key to boost the sensitivity of quantum sensors in trapped atom interferometers \cite{Zhang:2016c,Xu:2019b}, to realize fundamental test of quantum superposition states \cite{Robens:2015}, and to realize fault-tolerant quantum memories \cite{Briegel:2000}.

\begin{acknowledgments}
We are indebted to Gal Ness, Yoav Sagi, Yutaka Shikano, Gonzalo Muga, Eugene Sherman, Jacob Sherson, and Carrie Ann Weidner for insightful discussions.
This work is supported by the SFB/TR 185 OSCAR of the Deutsche Forschungsgemeinschaft (DFG).
A.A.~acknowledges support from the Reinhard Frank Foundation;
A.N.~acknowledges support from the Cluster of Excellence `CUI: Advanced Imaging of Matter' of the DFG, C.R.~is supported by a postdoctoral fellowship from the DFG.
\end{acknowledgments}

\bibliographystyle{apsrev4-1}
\bibliography{QBS_literature}

\begin{thebibliography}{64}
\bibitem [{non()}]{nonlocal}%
  \BibitemOpen
  \bibinfo {note} {By direct local coupling, we refer to Hamiltonians not
  containing terms of the kind $\ket{x+d}\kern-2pt\bra{x}$, displacing the
  system by a finite amount $d$.}\BibitemShut {Stop}%
\bibitem [{Fub()}]{FubiniStudy}%
  \BibitemOpen
  \bibinfo {note} {The Fubini-Study metric is the only Riemannian metric on the
  projective Hilbert space that, up to a constant factor, is invariant under
  all unitary transformations, thus including any possible coherent time
  evolution.}\BibitemShut {Stop}%
\bibitem [{qua()}]{quasiparticles}%
  \BibitemOpen
  \bibinfo {note} {In the related problem of transporting a quasiparticle
  (e.g., transferring an atomic excitation via an optical resonator), nonlocal
  Rabi coupling between initial and final states is possible, once the fast
  degrees of freedom (e.g., propagation of light) are adiabatically eliminated.
  For the transport of a massive particle, such an effectively nonlocal
  coupling fundamentally does not exist.}\BibitemShut {Stop}%
\end{thebibliography}

\clearpage

\makeatletter
\let\@hangfrom@section@orig=\@hangfrom@section
\let\@hangfroms@section@orig=\@hangfroms@section
\let\@sectioncntformat@orig=\@sectioncntformat

{\def\@addtoreset#1#2{}%
\appendix}
\@addtoreset{figure}{chapter}

\setcounter{equation}\z@%
\setcounter{figure}\z@
\def\thefigure{S\@arabic\c@figure}
\def\theequation@prefix{S}%

\let\section=\old@Section@Cmd

\let\@hangfrom@section=\@hangfrom@section@orig
\let\@hangfroms@section=\@hangfroms@section@orig
\let\@sectioncntformat=\@sectioncntformat@orig

\def\thesection{S\arabic{section}}

\makeatother

\makeatletter
\def\titleblock@produce{\begingroup \let\old@title=\@title\def\@title{Supplemental Material: \old@title} \ltx@footnote@pop \def \@mpfn {mpfootnote}\def \thempfn {\thempfootnote }\c@mpfootnote \z@ \let \@makefnmark \frontmatter@makefnmark \frontmatter@setup \thispagestyle {titlepage}\frontmatter@title@produce\groupauthors@sw {\frontmatter@author@produce@group }{\frontmatter@author@produce@script }\frontmatter@RRAPformat {\expandafter \produce@RRAP \expandafter {\@date }\expandafter \produce@RRAP \expandafter {\@received }\expandafter \produce@RRAP \expandafter {\@revised }\expandafter \produce@RRAP \expandafter {\@accepted }\expandafter \produce@RRAP \expandafter {\@published }}\@ifx@empty \@pacs {}{\@pacs@produce \@pacs }\@ifx@empty \@keywords {}{\@keywords@produce \@keywords }\par \frontmatter@finalspace \let\@title=\old@title\endgroup}

\makeatother

\originalmaketitle

\section{Atom trapping and cooling}
\label{app:ExperimentalApparatus}

We load ${}^{133}$Cs atoms from the background gas into a magneto-optical trap (MOT) and subsequently transfer them into a superimposed one-dimensional optical lattice with a trap depth $U_0\approx k_B \times \SI{400}{\micro\kelvin} \approx 4000\,E_\text{rec}$, where $k_B$ is the Boltzmann constant and $E_\text{rec} = (\hbar k)^2/(2m)=2\pi\hbar\times\SI{2}{\kilo\hertz}$ is the recoil energy; here, $k=2\pi/\lambda$ is the wavenumber associated with the wavelength $\lambda$ of the optical lattice, $m$ is the mass of cesium atoms, and $\hbar$ is the reduced Planck constant. The initial number of atoms is obtained by fluorescence imaging under near-resonant molasses illumination with an exposure time of $\SI{400}{\milli\second}$. A typical sample consists of \num{30} atoms loaded sparsely over \num{100} lattice sites. The molasses also cools the atoms further down by polarization gradient cooling. Adiabatically lowering the lattice trap depth to $k_B\times\SI{80}{\micro\kelvin} \approx 800E_\text{rec}$ further cools the atoms down to around $\SI{10}{\micro\kelvin}$. This temperature corresponds to a longitudinal ground state population of around $\SI{40}{\percent}$ as determined by microwave sideband spectroscopy.

A weak magnetic field of \SI{3}{\gauss} along the lattice axis provides a well-defined quantization axis.  Relative to the quantization axis, we select two hyperfine states of the ground state for the atom transport experiments, $\ket{\uparrow} = \ket{F=4,m_F = 4}$ and $\ket{\downarrow} = \ket{F=3,m_F = 3}$. In interferometer transport experiments, we use a superposition of both states, while for the other transport experiments we use state $\ket{\uparrow}$.

We cool the atoms down to the vibrational ground state along the longitudinal lattice direction by resolved sideband cooling using microwave radiation at \SI{9.2}{\giga\hertz} \cite{Belmechri:2013}.
More specifically, microwave sideband cooling is achieved by driving the cooling sideband $\ket{\uparrow,n}$ to $\ket{\downarrow,n-1}$, thereby removing one vibrational energy quantum $\hbar\,\omega_\text{HO}$, while simultaneously repumping the atoms to $\ket{\uparrow}$; here $\omega_\text{HO}=2\pi$ denotes the harmonic oscillation frequency,
\begin{equation}
    \omega_\text{HO} = 2\pi/\tau_\text{HO} =  2\pi \sqrt{\frac{2U_0}{m\lambda^2}}.
\end{equation}
Microwave sideband transitions are enabled by displacing one of the spin potentials by around \SI{17}{\nano\meter} along the lattice axis, lifting the orthogonality between different vibrational states. After sideband cooling for $\SI{20}{\milli\second}$, an longitudinal ground state population of typically \SI{96}{\percent} is reached.

In order to reduce the transverse temperature of the atoms, during molasses cooling we superimpose to the optical lattice a blue-detuned donut-shaped beam. Thereby, we increase the confinement of the atoms in the direction transverse to the optical lattice. By subsequently ramping down the intensity of the donut beam adiabatically, we lower the transverse temperature to $T_\perp\approx \SI{1}{\micro\kelvin}$.  

\section{Spin-dependent optical lattices}
\label{app:lattice_potential}

The optical lattice is operated at  $\lambda=\SI{865.9}{\nano\meter}$, a so-called ``magic'' wavelength allowing atoms in the state $\ket{\uparrow}$ to be trapped only by the right-handed circularly polarized (R-polarized) light, while atoms in the  state $\ket{\downarrow}$ are predominantly trapped by the left-handed circularly polarized (L-polarized) light. The dipole trap potentials for the two spin states are
\begin{subequations}
\begin{eqnarray}
U_{\uparrow} &=& -\alpha I_R \\
U_{\downarrow} &=& -\alpha \left(\frac{7}{8} I_L + \frac{1}{8} I_R\right).
\label{eq:7818}
\end{eqnarray}
\end{subequations}
where $I_R$ and $I_L$ denote the intensity of the two circular polarization components of the lattice laser field, and the proportionality constant $\alpha$ only depends on cesium polarizability.

\begin{figure}
	\centering
	\includegraphics[width=1\columnwidth]{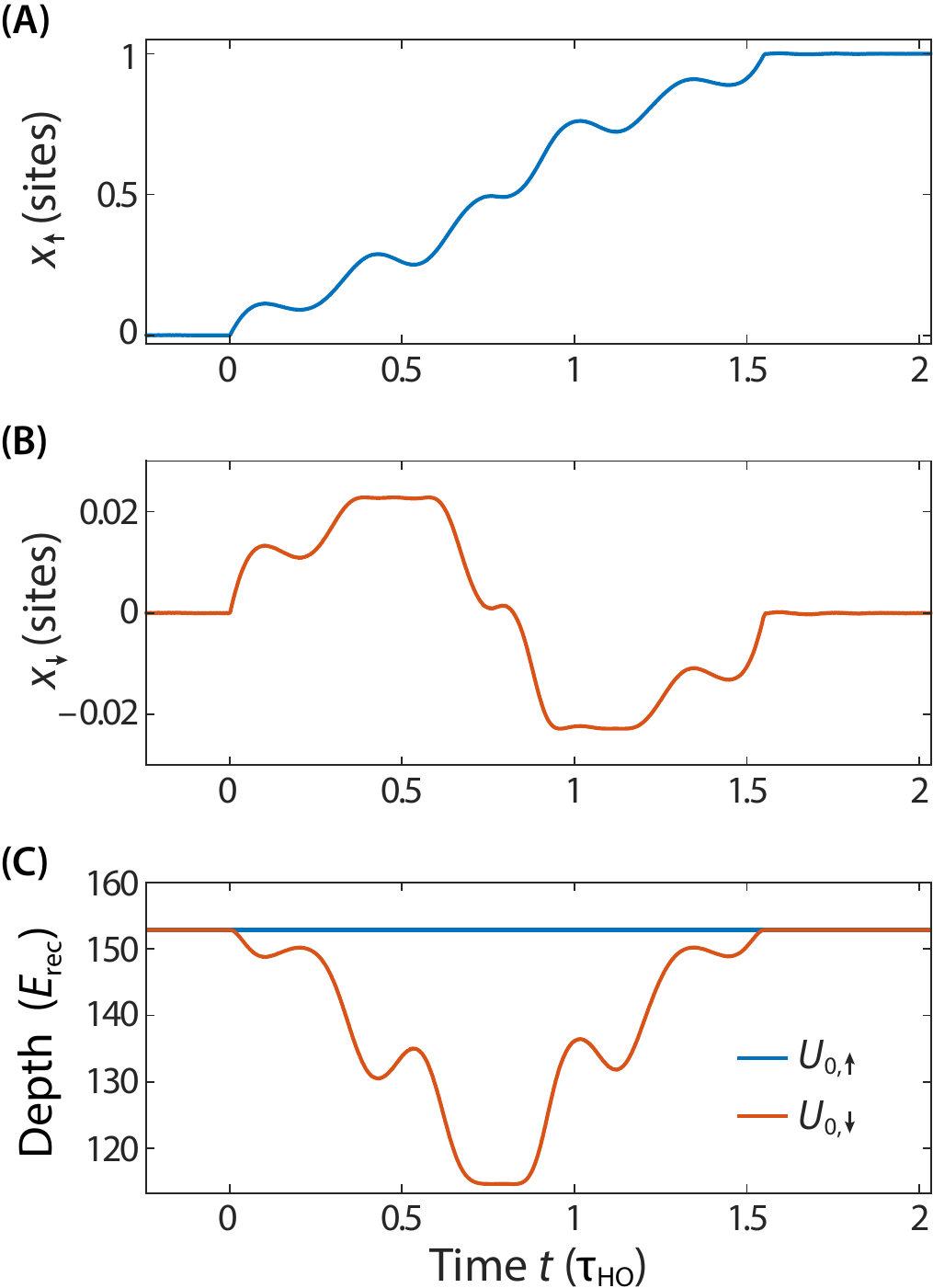}
	\caption{\textbf{Cross-talk between spin-dependent potentials.} 
    Example of transport ramp as in Fig.~\figref{Fig:Transport_Ramp}{d} when varying $x_R(t)$ without compensating $x_L(t)$, i.e., $x_L$$(t)=x_L(0)$. (A) The position of the spin-up potential only depends on the R-polarized standing wave, $x_\uparrow(t)=x_R(t)$. The position (B) and depth (C) of the spin-down potential are modulated because of the cross-talk contribution from the moving R-polarized standing wave, see \eqref{eq:latticeCrosstalkUdown}.}
	\label{FigApp:latticeCrosstalk}
\end{figure}

In order to create two fully independent optical conveyor belts transporting atoms selectively in either one of the two spin states, we employ a polarization-synthesized beam, where the phases $\phi_R$ and $\phi_L$ and the amplitudes of its left- and right-handed circularly polarized components are steered with high precision \cite{Robens:2018}.  By interfering the polarization-synthesized beam with a counter-propagating reference beam of fixed linear polarization, we create two perfectly superposed standing waves. The position of each standing wave
\begin{equation}
x_{R,L}(t)  = \frac{\lambda}{2}\frac{\phi_{R,L}(t)-\phi_0}{2\pi} 
\label{eq:latticePosition}
\end{equation}
is independently controlled by the  phase $\phi_{R,L}(t)$ relative to the phase $\phi_0$ of the counter-propagating reference beam. The conveyor belt potential for an atom in state $\ket{\uparrow}$ is simply
\begin{equation}
	\label{eq:U_up_arrow}
U_\uparrow(x,t) = - U_{0,\uparrow} \cos^2\left\{k\left[x-x_\uparrow\left(t\right)\right]\right\},
\end{equation}
with $x_\uparrow{\,=\,}x_R$ and $U_{0,\uparrow} {\,=\,} \alpha\, I_R>0$ being the trap depth; for the sake of notation, we simply use $U_0$ to refer to $U_{0,\uparrow}$ when only state $\ket{\uparrow}$ is involved.
Whereas, because both polarization components contribute to $U_\downarrow$ in Eq.~(\ref{eq:7818}), the conveyor belt potential for an atom in state $\ket{\downarrow}$ takes the form
\begin{equation}
	\label{eq:U_down_arrow}
U_\downarrow(x,t) = -U_{\text{offs},\downarrow} - U_{0,\downarrow} \cos^2\left\{k\left[x-x_\downarrow\left(t\right)\right]\right\},
\end{equation}
with
{\makeatletter
\@centering=-25.0pt plus 1000.0pt%
\makeatother
\begin{subequations}
\begin{eqnarray}
U_{0,\downarrow} &=& \frac{\alpha}{8} \sqrt{I_L^2+49I_R^2+14 I_L I_R \cos(\phi_R-\phi_L)}, \\
U_{\text{offs},\downarrow} &=& \frac{\alpha}{16}( I_L + 7I_R)-\frac{1}{2}U_{0,\downarrow},\\
x_\downarrow &=& \frac{\lambda}{4\pi}\arctan\left(\frac{I_L \sin(\phi_L)+7I_R\sin(\phi_R)}{I_L \cos(\phi_L)+7I_R\cos(\phi_R)}\right).
\end{eqnarray}%
\label{eq:latticeCrosstalkUdown}%
\end{subequations}%
}%
Here, $U_{0,\downarrow}>0$ and $U_{\text{offs},\downarrow}>0$ are the contrast and offset of the spin-down conveyor belt potential (see Fig.~\figref{FigApp:latticeCrosstalk}{}).

The phase of each of the two polarization components $\phi_{R,L}(t)$ is controlled by two independent optical phase-locked loops (OPLLs) with respect to a common reference beam, using two acousto-optical modulators as actuators. The set-points of the OPLLs are controlled by a direct digital frequency synthesizer (AD9954 by Analog Devices), enabling fast pre-programmed arbitrary phase ramps.
The control system has a bandwidth of $\SI{800}{\kilo\hertz}$ and a slew rate of $\SI{0.84}{\radian \per \micro\second}$, equivalent to \num{0.13} lattice sites per $\SI{}{\micro\second}$.

During the atom interferometer sequence described in the main text, the spin-down conveyor belt is kept static in order to preserve the spin-down wave function as a reference.
To that purpose, we actively compensate the effect of the moving R-polarized standing wave onto $U_\downarrow$ during the transport of the spin-up potential.
We therefore  suppress the position modulation with a compensation ramp $\phi_L$ (blue trajectory in the inset of Fig.~\ref{Fig:Interferometer}{} of the main text) that maintains $x_\downarrow$ constant,
\begin{equation}
	\phi_L = -\arcsin\left(\frac{\phi_R}{7f(\phi_R)}\right), \label{eq:crosstalkCorrection}
\end{equation}
where $f(\phi_R)$ is a rather involved analytical expression depending on $\phi_R$.
We do not compensate the depth modulation $U_{0,\downarrow}$, Fig.~\figref{FigApp:latticeCrosstalk}{c}, because motional excitations of atoms in state $\ket{\downarrow}$ are predominantly caused by position modulation $x_\downarrow$, when this latter is not properly compensated.

\section{Simulations of atom transport}
\label{app:simulations}

For the numerical simulations of atom transport, we consider a one-dimensional model of the conveyor belt potential, as introduced in Sec.~\ref{app:lattice_potential}, corresponding to the  Hamiltonian
\begin{equation}
    \hat{H}(t)  = \frac{\hat{p}^2}{2m} + U_0 \cos^2\{k[\hat{x} - x_\text{trap}(t)]\}.
\label{eq:Hamiltonian}
\end{equation}
We assume that the atom occupies initially the lowest energy state of $\hat{H}(0)$.
We compute the evolution of the wave packet in discrete time steps using the Strang split-step integration method \cite{MacNamara2016}.

In the transverse directions, a small, but nonzero temperature $T_\perp$ characterizes the initial state of the atoms, see Sec.~\ref{app:ExperimentalApparatus}.
For the atom transport problem, the motion of atoms in the transverse directions can be considered as frozen.
This assumption is justified by the large separation between the time scales of the longitudinal ($\SI{20}{\micro\second}$) and transverse (\SI{1}{\milli\second}) motion.
However, because of the thermal distribution of transverse positions, atoms experience a different trap depth $U_0$ depending on their distance from the lattice axis (inhomogeneous broadening).
Such a distribution of trap depths reduces the transport fidelity, especially for short transport durations close to the quantum speed limit, see Fig.~\ref{Fig:QSL_measurement} of the main text.
In the numerical simulations, 
we take into account the thermal distribution of transverse positions by assuming a two-dimensional Boltzmann distribution in the harmonic approximation of the transverse energy potential \cite{Belmechri:2013}
\begin{equation}
    \mathcal{P}(r,T_\perp)=\frac{m \hspace{0.5pt} \omega_\perp^2}{k_B T_\perp}\hspace{1pt}r\exp\left(-\frac{m \hspace{0.5pt} \omega_\perp^2r^2}{2 \hspace{0.3pt} k_B T_\perp}\right),
\end{equation}
where $r$ is the transverse distance from the lattice axis and $\omega_\perp$ is the transverse trap frequency. The effective trap depth experienced by atoms as a function of $r$ is
\begin{equation}
    U_0(r) = U_0(0) \exp\left( -\frac{2r^2}{w_{\text{DT}}^2} \right),
\end{equation}
where $w_{\text{DT}}$ is the lattice beam waist and $U_0(0)$ is the depth on the lattice axis. The average fidelity for a thermal ensemble of atoms is then given by 
\begin{equation}
    \mathcal{F}(\tau,T_\perp) \vspace{-2pt} = \hspace{-1pt} \int_0^{\infty}\hspace{-4pt}\mathrm{d}r \hspace{2pt} \mathcal{F}(U_0(r)) \,\mathcal{P}(r,T_\perp),
    \label{eq:EffectiveFidelityOfThermalAtoms}
\end{equation}
where
\begin{equation}
	\mathcal{F}(U_0) = \big|\hspace{-2pt}\braOpket{\psi_\text{target}}{\hat{V}\hspace{-0.3pt}(\tau,U_0)}{\psi_\text{init}}\hspace{-2pt}\big|^2\hspace{1pt}.
\end{equation}
Here, $\hat{V}\hspace{-0.3pt}(\tau,U_0)$ denotes the operator evolving the state for a time $\tau$ according to the Hamiltonian in \eqref{eq:Hamiltonian} with a trap depth $U_0$.
In practice, the integral in \eqref{eq:EffectiveFidelityOfThermalAtoms} is replaced by a trapezoidal sum over about \num{10} different discrete values of $r$.

\section{Precision optical measurement of transport ramps}
\label{app:opticalMeasurement}

Measuring the actual trajectory of the conveyor belt with high precision is important to achieve high fidelity transport operations.
Indeed, knowledge of the actual trajectory allows us to compensate for deviations from the target optimal trajectory  $x_\text{trap}(t)$; see Sec.~\ref{app:deconvolution}.

To that purpose, we developed an interferometric technique to reconstruct in situ the trajectory $x_\uparrow(t)$ and $x_\downarrow(t)$ of the optical conveyor belts for the two spin states:
The conveyor belt trajectories are inferred via Eqs.~(\ref{eq:U_up_arrow}) and (\ref{eq:U_down_arrow}) from the positions $x_{R}(t)$ and $x_{L}(t)$ of the R- and L-polarized optical standing waves, which are in turn obtained via \eqref{eq:latticePosition} from a time-resolved measurement of the optical phases $\phi_R(t)$ and $\phi_L(t)$ of the R- and L-polarized components that form the polarization-synthesized beam of the spin-dependent optical lattice.

\begin{figure}
	\centering
	\includegraphics[width=1\columnwidth]{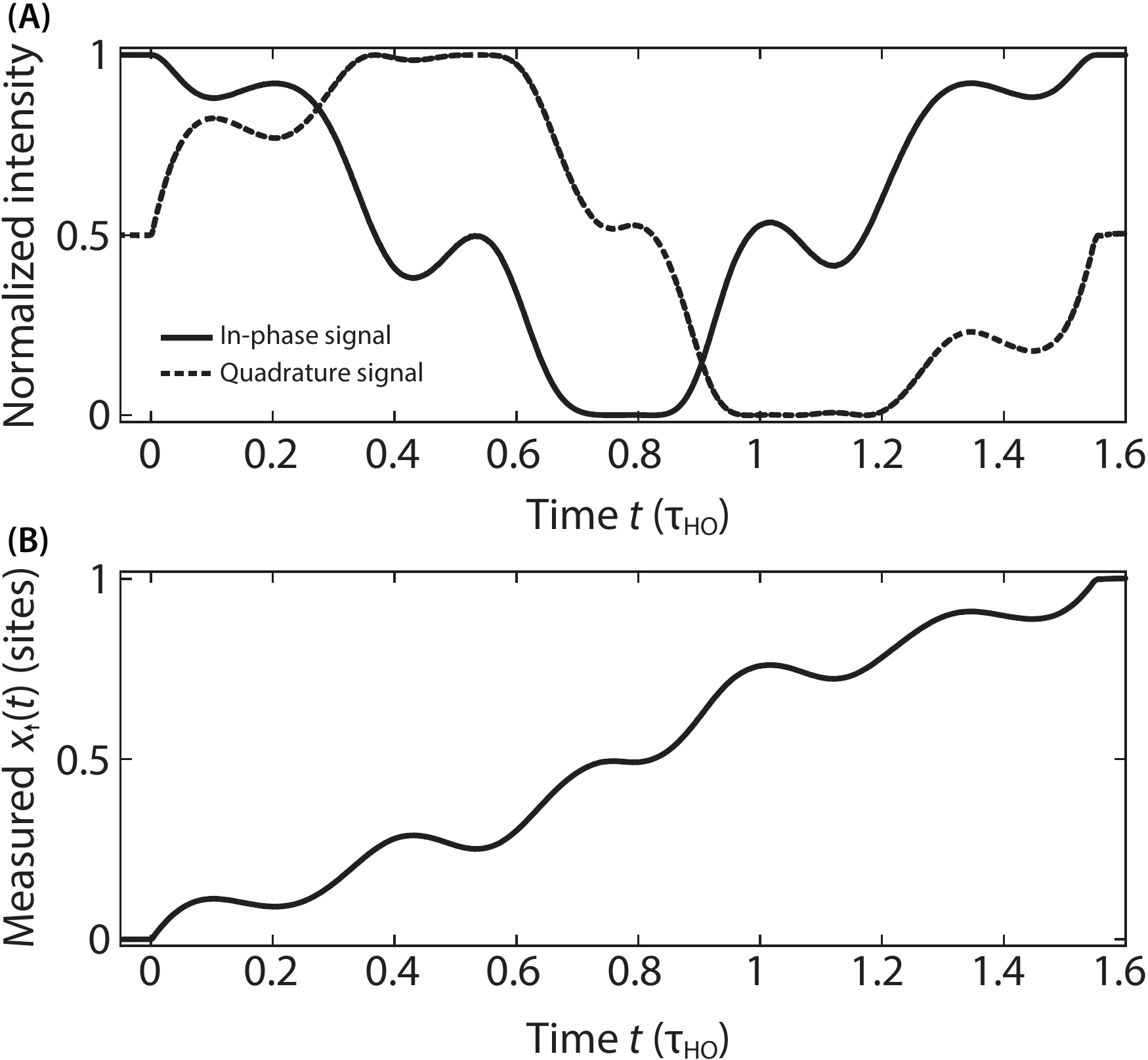}
	\caption{\textbf{Optical interferometric measurement of the trap trajectory.}
      	(A) Phase quadrature measurement of $\phi_R(t)$, showing the normalized intensities recorded after a Glan-Laser polarizer for the in-phase, $\{1+\cos[\phi_R(t)]\}/2$, and the quadrature signal, $\{1+\sin[\phi_R(t)]\}/2$.
	    (B) Displacement $x_\uparrow(t) = (\lambda/2)\,\phi_R(t)/(2\pi)$ of the optical conveyor belt, reconstructed from the in-phase and quadrature signals.
	 	 }
	\label{FigApp:OpticalMeasurementTRamp}
\end{figure}

The two phases are measured by using an optical phase quadrature detection scheme, which consists in inserting a Glan-Laser polarizer directly into the optical path of the polarization-synthesized beam, and detecting the intensity signal produced by the two interfering R- and L-polarized components.
If for example we aim to detect $\phi_{R}(t)$, we then hold $\phi_{L}(t)$ constant at either $\phi_{R}(0)$ or $\phi_{R}(0)+\pi/2$.
The recorded interference signals correspond to the in-phase and quadrature components of $\phi_{R}(t)$, respectively, from which it is straightforward to obtain $x_R(t)$; see Fig.~\ref{FigApp:OpticalMeasurementTRamp}.

\section{Avoiding distortions caused by bandwidth limitation}
\label{app:deconvolution}

Deviations from the target optimal trajectory, which are caused by the limited bandwidth of the control system, must be compensated in order to realize high fidelity transport operations.

\begin{figure}[b]
	\centering
	\includegraphics[width=1\columnwidth]{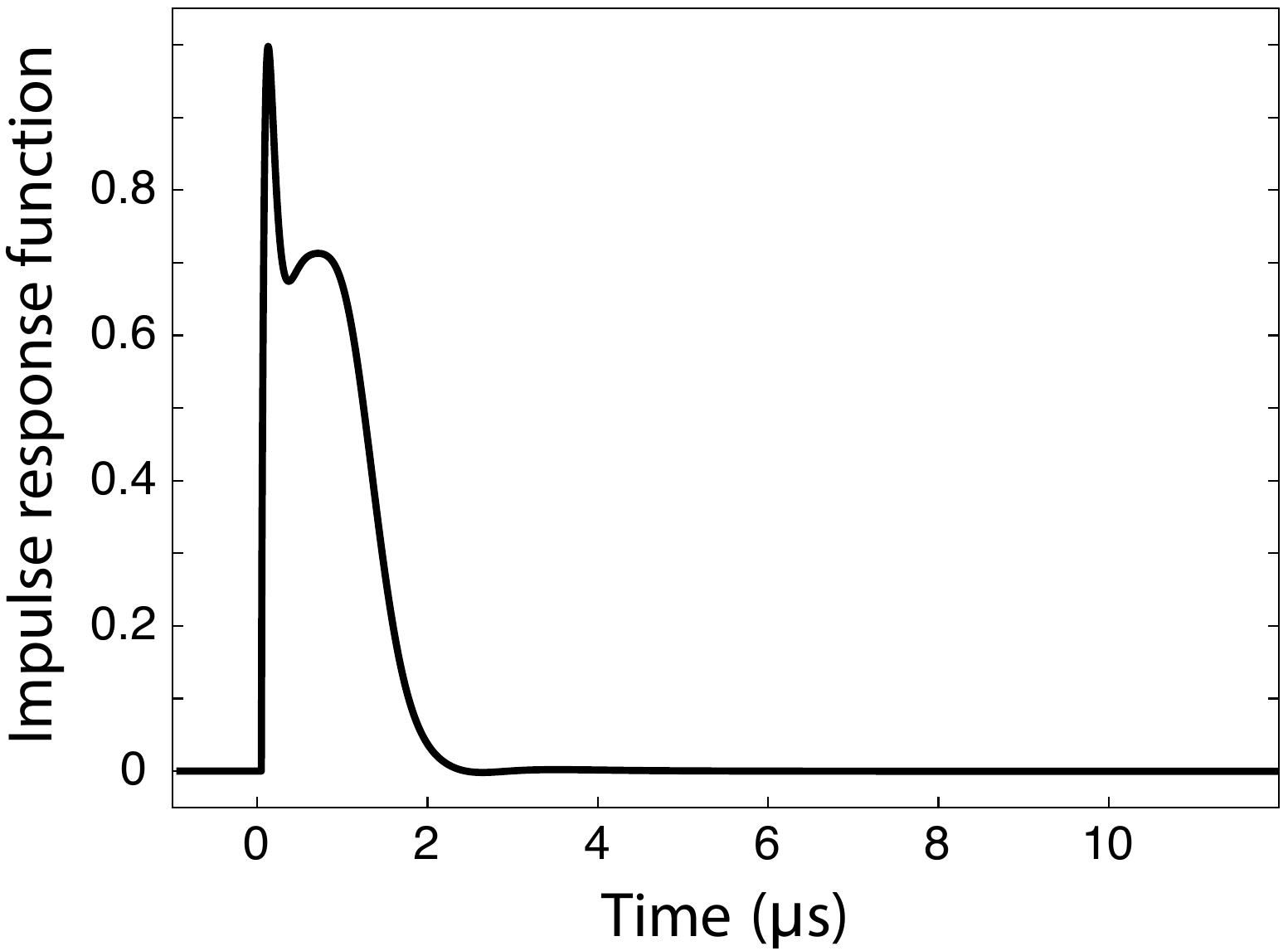}
	\caption{\textbf{Reconstructed impulse response function.} 
The control bandwidth is mainly limited \cite{Robens:2018} by a time delay, which originates from the acousto-optical modulators employed in the optical phase-locked loop to control the phases $\phi_{R,L}(t)$, see Sec.~\ref{app:ExperimentalApparatus}.
	 }
	\label{FigApp:IRF}
\end{figure}

To that purpose, we initially assume a linear time-invariant control system, implying that its response to an external drive is fully characterized by its impulse response function.
The impulse response function can be obtained as the derivative of the step response, 
which we measure with the technique described in Sec.~\ref{app:opticalMeasurement} by recording the actual position of the conveyor belt after driving a sudden, small step of its position.
The resulting impulse response function (Fig.~\ref{FigApp:IRF}) extends over a couple of microseconds, limiting the control bandwidth to below $\SI{1}{\mega\hertz}$.
By deconvolving the target optimal trajectory $x_\text{trap}(t)$ with the impulse response function, we obtain a first approximation of the external drive signal to be applied in order to avoid signal distortions.

In a second step, in order to also take into account non-linearities of the control system, we iteratively reduce 
 the residual deviations between the actual and the optimal target trajectory.
In each iteration, the residual deviations are measured with the technique in Sec.~\ref{app:opticalMeasurement}, and a fraction of them (typically 0.4 to avoid instabilities)
is subtracted from $x_\text{trap}(t)$ before deconvolution.
After 10 iterations, the difference between the measured and the target trajectories are below $\SI{2}{\percent}$ of a lattice site over the whole transport duration.
An example can be seen in Fig.~\figref{Fig:Transport_Ramp}{d} of the main text.

\section{Precision Measurement of transport fidelity}
\label{app:detection_fidelity}

The fidelity $\mathcal{F}$ of a transport operation is given by the fraction of atoms occupying the motional ground state $\ket{\psi_\text{target}}$ of the conveyor belt potential at the target position, as defined in \eqref{eq:fidelity_def} of the main text.

We measure the ground state fraction with a detection scheme that selectively removes atoms in higher motional states from the trap while retaining those in the ground state \cite{Belmechri:2013}:
All atoms are first transferred from $\ket{\psi_\uparrow}$ to $\ket{\psi_\downarrow}$ with a fast microwave $\pi$-pulse on the carrier transition, $\ket{\uparrow,n}\xrightarrow{}\ket{\downarrow,n}$.
Subsequently, the relative position $x_\uparrow(t){\,-\,}x_\downarrow(t)$ between the spin-up and spin-down conveyor belts is adiabatically increased from zero to around \SI{17}{\nano\meter} in order to enable microwave transitions on the motional sidebands.
We perform \num{10} repetitions of a removal cycle, where first a microwave pulse on the sideband $\ket{\downarrow,n}\xrightarrow{}\ket{\uparrow,n-1}$ transfers all atoms, except those in the ground state, to the spin-up state, and then a push-out beam resonant to the transition $\ket{F=4}\xrightarrow{}\ket{F'=5}$ removes the transferred atoms by radiation pressure.
The remaining fraction of atoms indicates the motional ground state population, with a typical statistical uncertainty at the \SI{2}{\percent} level.

To compensate for the imperfect initial state preparation, the reported values of the transport fidelity $\mathcal{F}$ are normalized by the fidelity of the initial state preparation (around \SI{96}{\percent}, see Sec.~\ref{app:ExperimentalApparatus}), which is measured by the same technique, but omitting the transport operation.

The fraction of atoms in the motional ground state, as measured with this scheme, does not discriminate whether the transported atom ends up in the ground state of the target site (true positive) or in that of an adjacent site of the optical lattice (false positive).
The latter possibility has however a negligible probability to occur, unless the transport duration is significantly shorter than the quantum brachistochrone time $\tau_\text{QB}$; see Fig.~\ref{Fig:QSL_measurement} of the main text.
Such false positive events could be separately detected and filtered out by resolving the individual lattice sites \cite{Alberti:2016} in addition to measuring the ground state probability.

\section{Ansatz for optimal transport trajectories}
\label{app:ClassicalLimitAndInitialGuess}

\begin{figure}
	\centering
	\includegraphics[width=1\columnwidth]{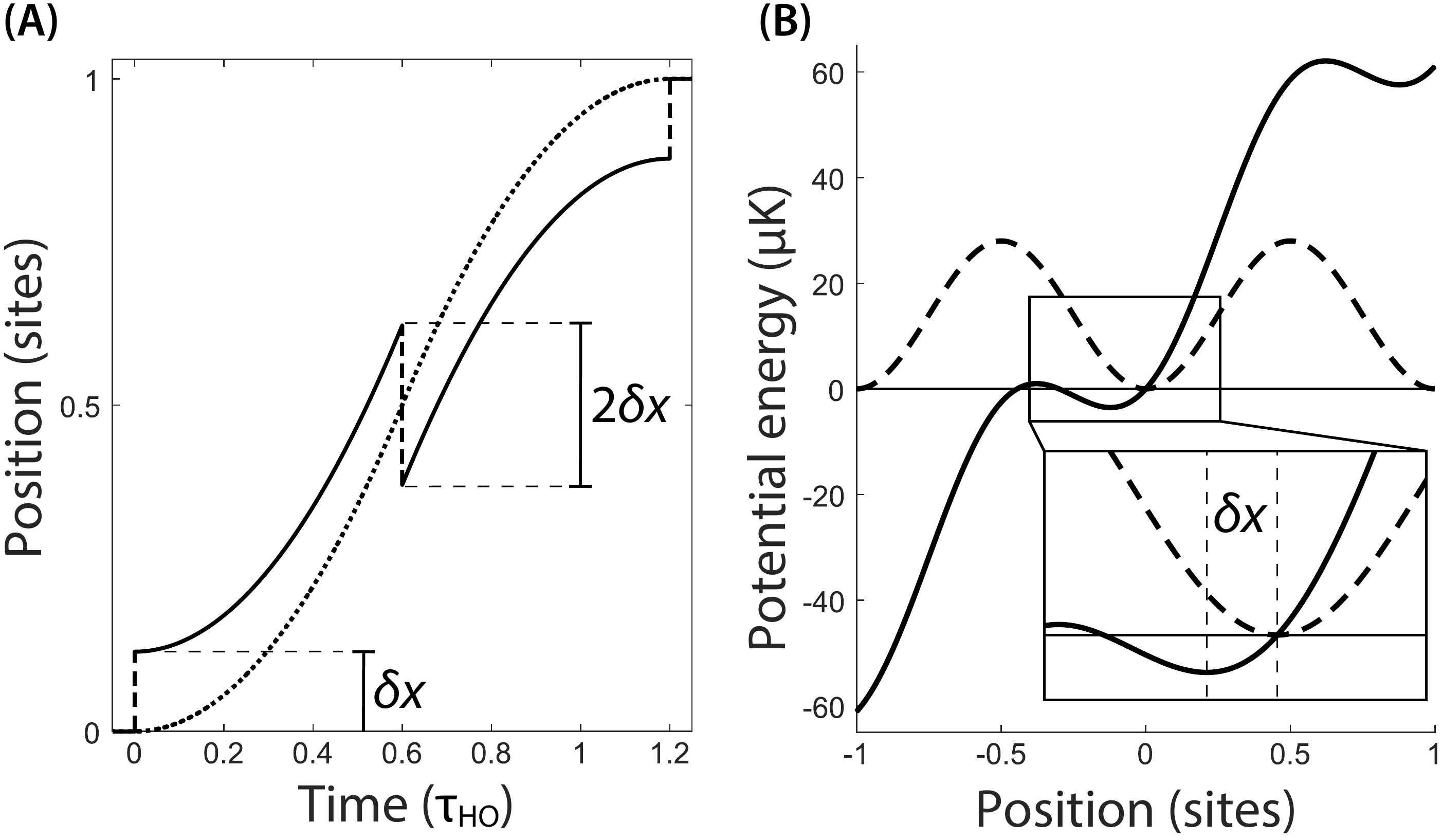}
	\caption{\label{FigApp:ClassicalTrajAndTiltedLattice}
	\textbf{Excitation-free classical trajectories.}
	 (A) The dotted curve shows the trajectory of a classical point particle being first constantly accelerated and then decelerated.
	 	 The solid curve is the trajectory of the trap $x_\text{ansatz}(t)$, see \eqref{eq:classicalTrajectory}, required to drive the particle along the dotted curve.
	 The example refers to $\tau=1.2\,\tau_\text{HO}$ and $d=\lambda/2$, whereas $\delta x$ is given by \eqref{eq:shiftOfInitialGuessTransport}.
	 (B) The potential in the non-inertial comoving frame (solid curve) is equal to the static potential (dashed curve) plus a linear tilt with slope $m\,\ddot{x}_\text{ansatz}(t)$.
	 The sudden shift by $\delta x$ keeps the particle at the position of the potential minimum in the comoving frame, avoiding motional excitations (e.g., slosh motion).
    	}
\end{figure}

For optimal control of the transport process (see Sec.~\ref{app:OptimalControlOfAtomTransport}), it is important to start with a good ansatz of the transport trajectory $x_\text{trap}(t)$. 
We obtain it considering the classical analog of the atom transport problem:
A classical point particle of mass $m$, initially at rest in a sinusoidal potential with lattice constant $\lambda/2$ and depth $U_0$, is to be transported over a distance $d$ in the shortest possible time such that it is again at rest after the transport.
The optimal strategy evidently is to maximally accelerate the particle during the first half of the transport and maximally decelerate it during the second half.
Thus, the optimal classical transport trajectory starts with a sudden lattice displacement equal to $\delta x = \lambda/8$, which places the particle at the point of steepest potential, where it is maximally accelerated.
The lattice potential is then moved together with the particle in order to maintain maximum acceleration until the particle reaches half of the transport distance.
At that point, the potential gradient is suddenly reversed by displacing the lattice by $-2\delta x$, thus ensuring maximum deceleration in the second half.
The particle reaches the target position at zero speed, where a final sudden displacement by $\delta x$ places the potential minimum at the particle's final position.
The duration of this process is the classical brachistochrone time,
\begin{equation}
	\tau_{\text{CB}}=\tau_\text{HO}\sqrt{2\,n/\pi},
\end{equation}
where $n=d/(\lambda/2)$ is the transport distance $d$ expressed in number of lattice sites.

This protocol can be extended to any transport duration $\tau\ge\tau_\text{CB}$ by reducing the constant acceleration and deceleration below the maximum value, yielding the trajectory (Fig.~\figref{FigApp:ClassicalTrajAndTiltedLattice}{a}) 
\begin{equation}
    x_\text{ansatz}(t) = \begin{cases}
		0 \quad &\text{for } t\le 0 \\
		\frac{d}{2}\left(\frac{t}{\tau/2}\right)^{\!2} + \delta x \quad &\text{for } 0<t<\tau/2 \\
		d-\frac{d}{2}\left(\frac{\tau-t}{\tau/2}\right)^{\!2} - \delta x \quad &\text{for } \tau/2<t<\tau \\
				d \quad &\text{for } t\ge \tau, \\
    \end{cases}
\label{eq:classicalTrajectory}
\end{equation}
with
\begin{equation}
    	\delta x = \frac{\lambda}{4\pi}\arcsin\!\left[\left(\frac{\tau_\text{CB}}{\tau} \right)^{\!2}\right]\leq \frac{\lambda}{8}.
\label{eq:shiftOfInitialGuessTransport}
\end{equation}
The effect of the sudden steps by $\delta x$ is best understood considering the dynamics from the reference frame comoving with the trap.
There, the classical particle is constantly kept at the position of the minimum of the tilted potential, thus avoiding in this reference frame motional excitations (e.g., slosh motion); see Fig.~\figref{FigApp:ClassicalTrajAndTiltedLattice}{b}.

We note that the trajectory $x_\text{ansatz}(t)$ resembles the transport trajectory proposed in Ref.~\cite{Zhang:2016}, which is obtained by minimizing the anharmonic contribution of the trap potential. This condition, in fact, can be shown to be related to minimizing the slosh motion as achieved by $x_\text{ansatz}(t)$.

\section{Optimal quantum control of transport trajectories}
\label{app:OptimalControlOfAtomTransport}

Optimal quantum control searches for the trajectory $x_\text{trap}(t)$ that maximizes the transport fidelity $\mathcal{F}(\tau,T_\perp)$ as defined in \eqref{eq:EffectiveFidelityOfThermalAtoms} for a given transport duration $\tau$ and transverse temperature $T_\perp$.
Relying  on numerical simulations of the transport problem (see Sec.~\ref{app:simulations}), we search for an optimal transport trajectory in the form of a Fourier series,
\begin{equation}
    x_\text{trap}(t) = d\,\frac{1-\cos(\nu_1t)}{2} + \sum_{j=1}^{j_\text{max}} b_j \sin(\nu_j t) , \quad t \in [0,\tau],
\label{eq:Fourier_Series}
\end{equation}
where the frequencies $\nu_j=\pi j/\tau$ are chosen to satisfy the boundary conditions $x_\text{trap}(0)=0$ and $x_\text{trap}(\tau)=d$.
We choose the maximum frequency \encapsulateMath{$\nu_{j_\text{max}}$} to lie within the bandwidth of our control system of around $\SI{800}{\kilo\hertz}$ to ensure that $x_\text{trap}$ can be faithfully reproduced in the transport experiments (see Sec.~\ref{app:deconvolution}).

Numerical simulations comparing the maximum fidelity reached by the optimization procedure as a function of the bandwidth of the control system show that the limitation to frequencies below $\nu_{j_\text{max}}$ has no significant effect in the range of parameters considered in this work.
In fact, because \encapsulateMath{$\nu_{j_\text{max}}$} is larger than $U_0/(2\pi\hbar)$, the control system bandwidth allows driving any relevant transition, i.e., any transition between pairs of discrete states of the trap, for which tunneling to neighboring sites is negligible.

Moreover, we conjecture that the optimal transport trajectory satisfies the point symmetry \encapsulateMath{$x_\text{trap}(t)=d-x_\text{trap}(\tau-t)$}, which is equivalent to reducing the search parameter space to the even Fourier coefficients $\{b_{2j}\}$ and thus taking $b_{2j+1}=0$.
This conjecture is supported by numerical studies, showing that when the search parameter space is unconstrained, the weight of the odd coefficients is negligible compared to that of the even coefficients.

For a robust convergence of the search algorithm to a global optimum of $\mathcal{F}(\tau,T_\perp)$, it is convenient to start the optimization procedure with good initial values of the coefficients $\{b_{2j}\}$ defining the transport trajectory.
For this purpose, based on physical intuition, we consider the trajectory defined in \eqref{eq:classicalTrajectory}, $x_\text{ansatz}(t)$, which is designed to avoid motional excitations of a classical point particle.
We project this ansatz into the form of \eqref{eq:Fourier_Series}, thus obtaining the initial set of control parameters $\{b_{2j}\}$ for the numerical optimization procedure.
We note here that alternative to $x_\text{ansatz}(t)$, one can choose as ansatz for the optimization procedure an optimal solution obtained for a slightly longer transport time \cite{Chen:2015b}.

While $x_\text{ansatz}(t)$ produces no motional excitations for a classical point particle, it does cause small, but not negligible wave packet deformations because of the anharmonicity of the potential.
These excitations, if not counteracted via optimal quantum control, would result in a loss of transport fidelity $\mathcal{F}(\tau)$, which becomes especially significant for $\tau$ close to $\tau_\text{QB}$.
Our numerical optimization of the transport process shows that optimal quantum control achieves this objective by avoiding too large motional excitations (e.g., breathing and slosh motion) in the reference frame comoving with the conveyor belt during the whole transport process.

Concerning the search algorithm, we use the interior-point method provided by MATLAB with the \texttt{fmincon} function, which allows us to include constraints.
We use constraints to limit the gradient of the trajectory to the maximum slew rate of the control system ($\SI{0.84}{\radian / \micro\second}$), which is determined by how fast the OPLL is able to track the change of its set-point; see Sec.~\ref{app:lattice_potential}.

We note that for transport over many lattice sites, more frequency components $\nu_j$ fit within the system bandwidth due to the longer transport time, resulting in a higher dimensional search parameter space.
In this case, using a reduced randomized basis of functions to represent $x_\text{trap}(t)$, as done by the DCRAB algorithm \cite{Rach:2015}, is expected to be preferable to exhaustively searching through the whole system bandwidth at once, as done here.

\section{Bang-bang control}
\label{app:bang-bang}

A widely used transport method is the so called bang-bang type of transport.
We here compare the fidelities achieved with our optimal control optimization procedure to the fidelities of two types of bang-bang transport protocols:
the linear transport and the parabolic transport.
For the linear transport, $x_\text{trap}(t)$ follows a trajectory with constant speed from the initial to the target position.
For the parabolic transport, $x_\text{trap}(t)$ is constantly accelerated with $\ddot{x}_\text{trap}(t)=a$ on the first half of the transport and constantly decelerated with $\ddot{x}_\text{trap}(t)=-a$ on the second half.

Both protocols are better understood in the reference frame comoving with the trap.
During the linear transport, the wave packet is subject to two momentum kicks:
one at the start and one at the end.
During the parabolic transport, the wave packet is subject to three position kicks:
one at the start by $-a/\omega$, one at half of the transport time by $2a/\omega$ and one at the end by $-a/\omega$.
In both cases, motional excitations are created after the initial kick.
However, the transport process can be timed so that the excitations created by the first and possibly middle kicks are undone by the last kick.
The simulated infidelities of the two transport types are shown in Fig.~\figref{FigApp:BangBangProtocols}{} and compared to the infidelity of the optimal control transport as well as the adiabatic transport discussed in Sec.~\ref{app:adiabatic_limit}.
The ``magic'' transport durations for which the transport brings the wave packet back to a minimally excited state lie close to multiples of approximately the harmonic period.
The small, but visible deviation from the harmonic period $\tau_\text{HO}$ can be understood to a very good approximation as the result of the anharmonic potential, which yields an effectively lower trap frequency $\tilde{\omega}_\text{HO} \approx \omega_\text{HO}-E_\text{rec}/\hbar$ and, correspondingly, an effectively longer oscillation period $\tilde{\tau}_\text{HO} \approx \tau_\text{HO} [1+E_\text{rec}\tau_\text{HO}/(2\pi\hbar)]$.

The dashed lines are the envelopes (worst-case infidelities) derived in the harmonic approximation for the two bang-bang protocols:
\begin{subequations}
	\begin{align}
		\label{eq:linear_tau_F}
		\tau_\text{linear}(\mathcal{F}) &= \tau_\text{HO}\frac{1}{\pi}\frac{\ell_\text{QGT}}{[-\log(\mathcal{F})]^{1/2}},\\
		\label{eq:bang_bang_tau_F}
		\tau_\text{parabolic}(\mathcal{F}) &= \tau_\text{HO}\frac{2}{\pi}\frac{\sqrt{\ell_\text{QGT}}}{[-\log(\mathcal{F})]^{1/4}}.
\end{align}
\end{subequations}

Their scaling with distance, $\ell_\text{QGT} \propto d$, indicates that the linear transport protocol is faster for short transport distances, whereas the parabolic transport is faster for long distances, since the trap can be accelerated to higher speeds.
Both are, however, much slower than the transport operation obtained by optimal control, which is also shown for comparison in Fig.~\figref{FigApp:BangBangProtocols}{}.

\section{Adiabatic limit}
\label{app:adiabatic_limit}

\begin{figure}
	\centering
	\includegraphics[width=1\columnwidth]{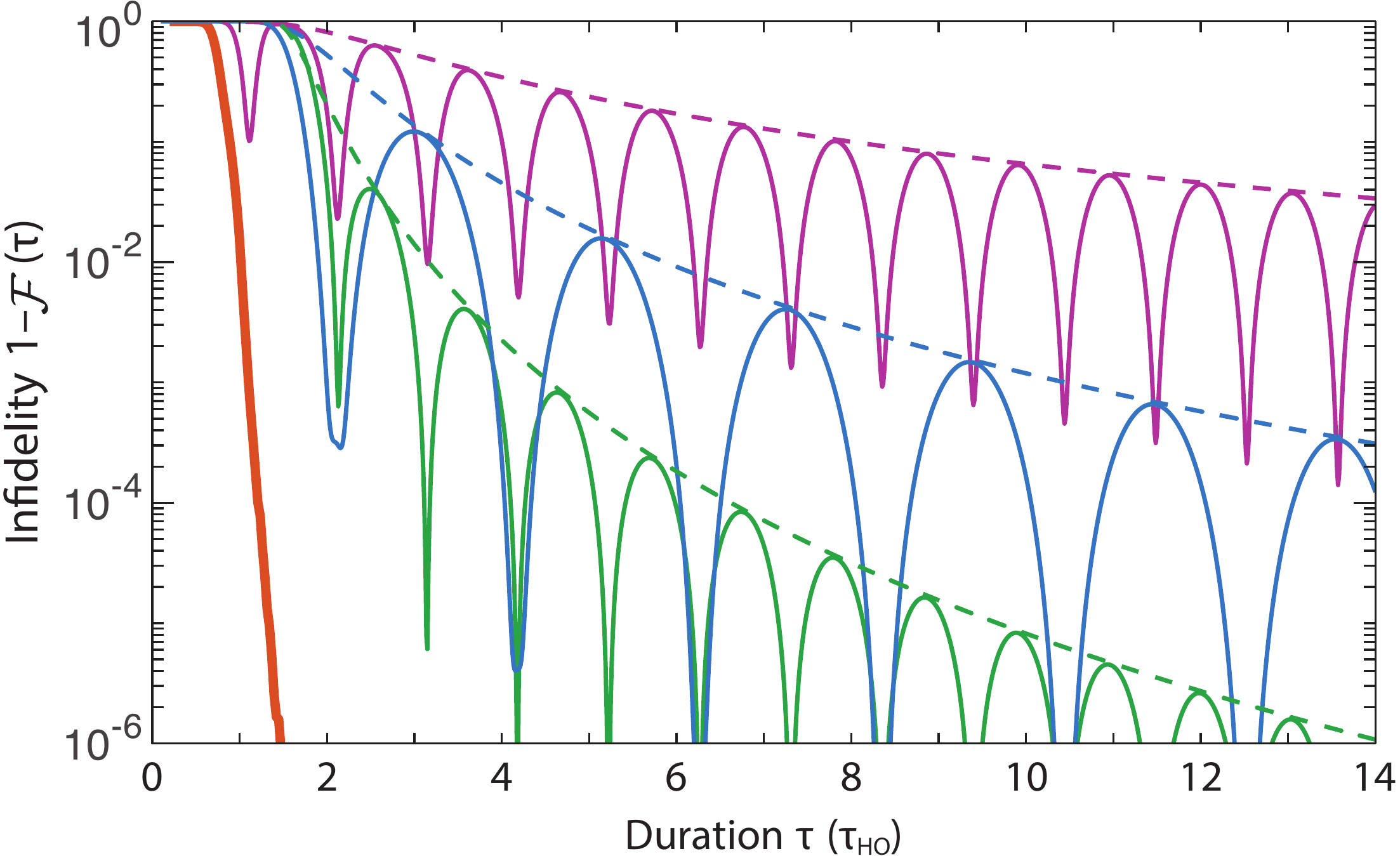}
	\caption{\textbf{Transport protocols compared.} 
	Infidelity computed numerically for the same conditions of Fig.~\ref{Fig:QSL_measurement} and $T_\perp = 0$ using different transport protocols. From top to bottom: a linear transport ramp (purple), parabolic control (blue), adiabatic control (green), optimal quantum control (thick red). The dashed lines represent the envelope functions according to Eqs.~(\ref{eq:linear_tau_F}), (\ref{eq:bang_bang_tau_F}), (\ref{eq:adiabatic_tau_F}).
	}
	\label{FigApp:BangBangProtocols}
\end{figure}

Adiabatic transport minimizes excitations of the wave packet during the entire transport by using smooth transport ramps.
As an example, we here consider ramps that follow a sinusoidal trajectory, which is continuous in position, velocity and acceleration,
\begin{equation}
    x_\text{trap}(t)=A\sin(2\pi t/\tau)+d\,t/\tau,
\end{equation}
where $A = -d/(2\pi)$ is chosen such that $\dot{x}_\text{trap}(0)=0$. For a given fidelity, we find in the harmonic approximation that the required worst-case duration of the adiabatic transport is
\begin{equation}
	\label{eq:adiabatic_tau_F}
	\tau_\text{adiabatic}(\mathcal{F}) = \tau_\text{HO}\sqrt{\frac{2}{3}+\left[\frac{\ell_\text{QGT}^2}{\pi^2(-\log(\mathcal{F}))}\right]^{1/3}}.
\end{equation}
This relation is shown as the dashed green curve in Fig.~\figref{FigApp:BangBangProtocols}{}.

\section{Estimation of geometric path length and energy spread}
\label{app:estimation_DeltaE}

A very good, analytic approximation of the  geometric path length can be obtained assuming that at the quantum speed limit the wave packet is steadily accelerated in the first half and steadily decelerated in the second half, meaning that the average position of the wave packet evolves as 
\begin{equation}
    \bar{x}_\text{QB}(t) \approx \begin{cases}
        2d\,(t/\tau)^2 \quad &\text{for } 0<t<\tau/2 \\
        -d+4dt/\tau-2d(t/\tau)^2 \quad &\text{for } \tau/2<t<\tau
    \end{cases}
\label{eq:semiclassicalTrajectory}
\end{equation}
Numerical simulations confirm that this assumption, where $\bar{x}_\text{QB}(t)$ is a smooth function of time,  is well fulfilled despite the much less regular shape of the optimal control transport trajectories $x_\text{trap}(t)$.
Moreover, we assume that quantum optimal control preserves the wave packet $\ket{\psi(t)}$ close to a coherent state $\ket{\alpha(t)}$, avoiding too large spreading and deformation, in particular when it approaches the points of steepest potential, where the trap loses its ability to confine.
The coherent state is specified by its phase space coordinates,
\begin{equation}
    \label{eq:alpha_definition}
    \alpha(t) = \frac{\bar{x}_\text{QB}(t)}{2\Delta x}+i\frac{m\,\dot{\bar{x}}_\text{QB}(t)}{2\Delta p},
\end{equation}
with the position and momentum width being $\Delta x \approx \sqrt{\hbar /(2m\omega)}$ and $\Delta p \approx \hbar/(2\Delta x)$.
Thus, the geometric path length $\ell_\text{QB}$, as defined in Eq.~(\ref{eq:AA_exp}) of the main text, is obtained by integrating the Fubini-Study differential form
\begin{equation}
\mathrm{d}s_\text{FS} = \left|\dot\alpha(t)\right|\mathrm{d}t,
\end{equation}
over the duration $\tau$. The integration produces
\begin{equation}
	\label{eq:app_approx_ellQB}
	\ell_\text{QB} \approx  \frac{d}{2\Delta x}\,f\!\left(\frac{\tau_\text{HO}}{\pi\tau_\text{QB}}\right),
\end{equation}
which corresponds to \eqref{eq:approx_ell} of the main text.

We note that the factor $d /(2\Delta x)$ can be identified with the geodesic length $\ell_\text{QGT}$ determined by the quantum geometric tensor~\cite{Kolodrubetz:2017,Bukov:2019}.
In the atom transport problem, the quantum geometric tensor $\chi_{\mu,\nu}$ reduces to a scalar quantity $\chi_{1,1}$ because of the single control parameter used to transport the atom, i.e., the conveyor belt position $x_\text{trap}$.
Its value is specified by the differential form $\mathrm{d}s_\text{QGT}^2 =$ \encapsulateMath{$\mathrm{d}x_\text{trap}\,\chi_{1,1}\,\mathrm{d}x_\text{trap}=1-\left|\braket{\psi_0(x_\text{trap}+ \mathrm{d} x_\text{trap})}{\psi_0(x_\text{trap})}\right|^2$},
where $\ket{\psi_0(x_\text{trap})} = \exp(-i \hat{p} \,x_\text{trap}/\hbar)\ket{\psi_\text{init}}$ denotes the ground state of the conveyor belt displaced to the position $x_\text{trap}$ ($\hat{p}$ is the momentum operator).
The physical meaning of the quantum geometric tensor is that of the Fubini-Study metric in the restricted manifold of states reachable by an adiabatic transformation of the control parameter $x_\text{trap}$.
An explicit computation of its value yields $\chi_{1,1} = (\Delta p/\hbar)^2$, from which we directly obtain
\begin{equation}
    \ell_\text{QGT} = \int_0^d \mathrm{d}s_\text{QGT} =\frac{d \,\Delta p}{\hbar} \approx \frac{d}{2\Delta x},
\end{equation}
where the last step
follows from the approximately Gaussian shape of the ground state.
Notably, the geodesic defined by the quantum geometric tensor, in stark contrast with the Fubini-Study geodesic, denotes a path that actual physical processes (e.g., adiabatic transformations) can follow.
In contrast to $\ell_\text{QB}$ in \eqref{eq:app_approx_ellQB}, $\ell_\text{QGT}$ is a purely geometric quantity independent of the out-of-equilibrium dynamics of the process.
Its length $\ell_\text{QGT}$ does however scale with the transport distance $d$, and it can be loosely interpreted as the number of local transformations necessary to carry out the transport process.	

For the determination of an upper bound on the energy spread $\Delta E$, we rely on the same assumptions made to estimate $\ell_\text{QB}$, i.e., an approximately coherent state evolving as specified in \eqref{eq:alpha_definition}.
A direct calculation of the instantaneous energy spread yields
\begin{multline}
	\label{eq:instant_delta_E}
	\Delta E(t) =\big[\braOpket{\psi(t)}{\hat H^2(t)}{\psi(t)}\! -\! \braOpket{\psi(t)}{\hat H(t)}{\psi(t)}^2\big]^{1/2}\\
	\approx \bigg[\dot{\bar{x}}_\text{QB}^2(t)\,\Delta p^2 +  \left(\frac{\partial U(x,t)}{\partial x}\right)^{\!\!2}\!\Delta x^2 \bigg]^{1/2}\!\!\!\!\big[1+\mathcal{O}(\eta^2)\big],
\end{multline}
where $U(x,t)$ refers to the lattice potential, as defined in Eqs.~(\ref{eq:U_up_arrow}) and (\ref{eq:U_down_arrow}), the derivative of the potential is computed at $x=\bar{x}_\text{QB}(t)$, and $\eta^2=E_\text{rec}/(\hbar \omega) = 1/\sqrt{4\,U_0/E_\text{rec}}$ is the Lamb-Dicke factor, which is negligible for the trap depths considered in this work.
The two terms in \eqref{eq:instant_delta_E} correspond to the leading contributions to the energy uncertainty,
\begin{subequations}
	\begin{eqnarray}
		\label{eq:DeltaK_t}
		\Delta K(t) &=& \Delta p\,|\dot{\bar{x}}_{\text{QB}}(t)|, \\
		\Delta U(t) &=& \Delta x\,\left|\frac{\partial U(x,t)}{\partial x}\right|_{x=\bar{x}_\text{QB}(t)},
	\end{eqnarray}
\end{subequations}
originating from the kinetic ($\Delta K$) and potential ($\Delta U$) energy.
The origin of the two terms can be intuitively understood if we consider the evolution of the wave packet in the reference frame comoving with $\bar{x}_\text{QB}(t)$. There, the wave packet is at rest and displaced from the center of the site by a distance $\bar{x}_\text{QB}(t){\,-\,}x_\text{trap}(t)$, where the potential has a nonvanishing slope $\partial U/\partial x$, which explains the potential contribution $\Delta U(t)$.
A Galilean transformation from the comoving to the laboratory reference frame introduces a term equal to $\dot{\bar{x}}_\text{QB}(t) \,\hat{p}$ to the Hamiltonian (Heisenberg representation), explaining the kinetic contribution $\Delta K(t)$.

We bound $\Delta U$ from above by replacing the derivative of the potential (i.e, the force applied to the wave packet) by its maximum value.  For the conveyor belt potentials in Eqs.~(\ref{eq:U_up_arrow}) and (\ref{eq:U_down_arrow}), the maximum of the derivative, $2\pi U_0/\lambda$, is reached at the positions $\pm\lambda/8$ relative to the center of the site, yielding the inequality
\begin{equation}
	\Delta E(t) < \big[\dot{\bar{x}}_\text{QB}^2(t)\,\Delta p^2 +  \left( 2\pi U_0/\lambda \right)^{2}\!\Delta x^2 \big]^{1/2}.
\end{equation}
Integrating this expression over time \cite{defDeltaH} gives an upper bound on the time-averaged energy uncertainty,
\begin{equation}
	\Delta E < \Delta E_\text{upper} =  \frac{\hbar}{\tau_\text{QB}}\,\ell_\text{QGT}\,f\!\left(\frac{\tau_\text{QB}}{2\,n\,\tau_\text{HO}}\right),
\end{equation}
which corresponds to Eq.~(\ref{eq:DeltaE_bound}) of the main text.

We note that at the quantum speed limit, for very long transport distances, $n=d/(\lambda/2)\gg1$, $\Delta E$ is dominated by the kinetic rather than the potential contribution,
\begin{equation}
	\label{eq:ratio_DeltaK_DeltaU}
	\frac{\Delta K}{\Delta U}  >  \frac{\Delta p}{\Delta x}\frac{d/\tau}{2\pi U_0/\lambda} = n\frac{\tau_\text{HO}}{\tau} \propto \sqrt{n}\gg 1,
\end{equation}
where $\Delta K$ and $\Delta U$ denote here the time average of $\Delta K(t)$ and $\Delta U(t)$, respectively; in this expression, the first inequality results from the foregoing upper bound on $\Delta U(t)$, whereas the proportionality assumption follows from the scaling $\tau\propto \tau_\text{HO} \sqrt{n}$ expected for a quantum brachistochrone process.
Hence, we find that in the limit of $n\gg1$, the energy uncertainty $\Delta E$ of a transport process at the quantum speed limit reduces to the time average of $\Delta K(t)$ in \eqref{eq:DeltaK_t},
\begin{equation}
	\label{eq:deltaE_long_distance}
	\Delta E \approx \frac{\hbar}{\tau_\text{QB}}\,\ell_\text{QGT}.
\end{equation}

\section{The Mandelstam-Tamm bound in the limit of long distances}
\label{app:scalingMandelstamTamm}

We investigate the scaling of the \MT bound, $\tau_\text{QB}\geq \tau_\text{MT}$, in the limit of long transport distances.
In its most general form \cite{Deffner:2017a}, when the initial $\ket{\psi_\text{init}}$ and target $\ket{\psi_\text{target}}$ states are not necessarily orthogonal, the \MT time reads
\begin{equation}
	\label{eq:general_MT}
	\tau_\text{QB}\geq\tau_\text{MT} = \frac{\ell_\text{geo}}{\Delta E/\hbar},
\end{equation}
where $\Delta E$ represents the time-averaged energy uncertainty \cite{defDeltaH}, and $\ell_\text{geo}$ denotes the geodesic length as measured by the Fubini-Study metric \cite{FubiniStudy}, $\ell_\text{geo} = \arccos(\left|\braket{\psi_\text{target}}{\psi_\text{init}} \right|)$.

Concerning the numerator in \eqref{eq:general_MT}, it is evident that $\ell_\text{geo}$ levels off to its maximum value, $\pi/2$, since for long distances, $d\gg \Delta x$, the target state is effectively orthogonal to the initial state.

Concerning the denominator in \eqref{eq:general_MT}, it can be shown, see \eqref{eq:deltaE_long_distance}, that for very long transport distances its expression is well approximated by
\begin{equation}
	\label{eq:delta_E_scaling}
	\Delta E/\hbar \approx \frac{d}{2\Delta x}\frac{1}{\tau_\text{QB}} \propto 
	\sqrt{d},
	\end{equation}
where the last step follows from the scaling $\tau_\text{QB} \propto \sqrt{d}$ expected for the quantum brachistochrone time $\tau_\text{QB}$ as a function of the transport distance $d$, see \eqref{eq:classical_bound} of the main text.
The scaling of $\Delta E$ in \eqref{eq:delta_E_scaling} results in the seemingly counter-intuitive fact that $\tau_\text{MT}$ is a monotonically decreasing function of the distance, in stark contrast with the monotonically increasing behavior of $\tau_\text{QB}$ with respect to the distance.

\end{document}